\documentclass[final,5p,times,twocolumn]{elsarticle}%
\newcommand{\tabfontsize}{\scriptsize}

\usepackage{amsmath,amssymb,amsfonts,mathtools}
\usepackage{amsthm} 
\usepackage{cases}

\newtheorem{definition}{Definition}[section]
\newtheorem{theorem}{Theorem}[section]
\newtheorem{lemma}{Lemma}[section]
\newtheorem{remark}{Remark}[section]

\usepackage{xcolor}
\usepackage{graphicx}                %
\usepackage[caption=false,font=normalsize]{subfig}  
\usepackage{booktabs}                
\usepackage{array,makecell,multirow} 
\usepackage{pifont}

\usepackage{algorithm}
\usepackage{algorithmic} 

\usepackage{url}
\usepackage[hidelinks]{hyperref}     %

\biboptions{sort&compress}

\DeclareMathOperator{\diag}{diag}
\DeclareMathOperator{\Diag}{Diag}

\def\bfx{\mathbf{x}}

\def\B{{\cal B}}
\def\K{{\cal K}}
\def\S{{\cal S}}
\def\T{{\cal T}}
\def\EDMAR{\mbox{EDMAR}}

\newcommand{\be}{\begin{equation}}
\newcommand{\ee}{\end{equation}}
\newcommand{\ba}{\begin{eqnarray}}
\newcommand{\ea}{\end{eqnarray}}
\newcommand{\bas}{\begin{eqnarray*}}
\newcommand{\eas}{\end{eqnarray*}}

\journal{Digital Signal Processing}
\begin{document}
\begin{frontmatter}



\title{A Robust EDM Optimization Approach for 3D Single-Source Localization with Angle and Range  Measurements}

\author[affBIT]{Mingyu Zhao}
\ead{3120225732@bit.edu.cn}

\author[affBIT]{Qingna Li\corref{cor1}}
\ead{qnl@bit.edu.cn}

\author[affPolyU]{Hou-Duo Qi}
\ead{houduo.qi@polyu.edu.hk}

\cortext[cor1]{Corresponding author.}

\affiliation[affBIT]{organization={School of Mathematics and Statistics, Beijing Institute of Technology},
  addressline={5 South Zhongguancun Street, Haidian District},
  city={Beijing},
  postcode={100081},
  state={Beijing},
  country={China}}

\affiliation[affPolyU]{organization={Department of Data Science and Artificial Intelligence, and Department of Applied Mathematics, The Hong Kong Polytechnic University},
  addressline={Hung Hom, Kowloon},
  city={Hong Kong},
  country={Hong Kong}}

\begin{abstract}
Accurate source localization in Multi-Platform Radar Networks (MPRNs) benefits from 
{jointly exploiting range and angle measurements, especially under noisy conditions}. 
In this paper, we propose a robust Euclidean distance matrix (EDM) optimization model 
{for 3D single-source localization (3DSSL), which integrates range measurements and angle information into a unified distance-based formulation and naturally supports the least absolute deviation ($\ell_1$-norm) criterion}.
{
In this formulation, the angle information is incorporated through lower and upper bounds on the source-to-sensor distances, which form box constraints in the EDM model.
This is achieved by reducing the 3D angle constraints to two-dimensional nonlinear optimization subproblems, whose global minimum and maximum values provide the required distance bounds.} 
To solve the resulting rank-constrained EDM problem, we develop an efficient algorithm based on the majorization penalty method. 
Extensive numerical experiments confirm that the proposed EDM model 
{outperforms leading vector-based solvers in localization accuracy while maintaining competitive computational efficiency}, particularly in low Signal-to-Noise Ratio (SNR) scenarios.
\end{abstract}



\begin{keyword}


Euclidean distance matrix optimization\sep Single-source localization\sep Multi-platform radar networks\sep Angle constraints\sep Range constraints\sep Penalty method
\end{keyword}

\end{frontmatter}



\section{Introduction}
\begin{figure}[t]
	\centering
		\includegraphics[width=0.9\linewidth]{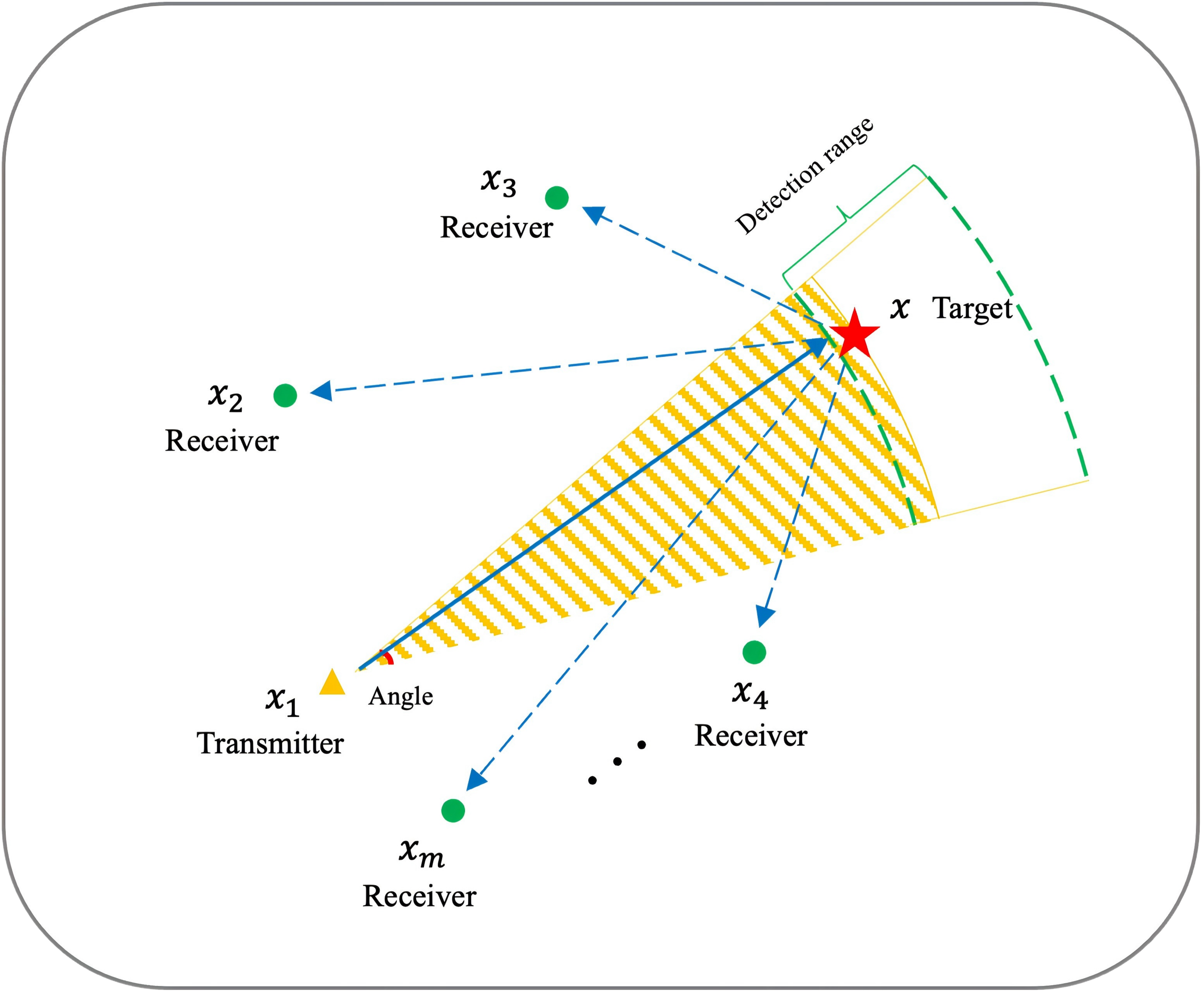}
	\caption{Illustration of SSLAR including one transmitter and multiple receivers.}
	\label{fig_intro}
\end{figure}

{Multi-platform radar networks (MPRNs) have attracted increasing attention as a promising sensing architecture for detection, localization, and tracking \citep{Re_MPRNs_3, Sun2022, Re_MPRNs_2, Re_Geng}. Compared with monostatic and bistatic radar systems, MPRNs offer significant advantages in terms of spatial diversity, wider coverage, and improved fault tolerance
\citep{Re_multibetter, zhang2020multistatic, Re_multibetter_1, wu2025multistatic}. In such systems, each radar node not only receives time-delay data, which can be converted into noisy distance measurements, but also benefits from additional angle and range constraints derived from the transmitter’s radiation pattern and detection range. At the same time, increasing the number of radar nodes leads to higher deployment, communication, and synchronization costs. Therefore, achieving high localization accuracy with a limited number of nodes requires the available angle and distance information to be exploited as effectively as possible. This is challenging because the resulting geometric formulation is highly nonconvex \citep{Re_Enhance_a}. In this paper, we consider the three-dimensional (3D) single-source localization problem with angle and range information (SSLAR) in MPRNs.} The interplay between angle and range constraints in SSLAR is illustrated in Fig.~\ref{fig_intro}.


\subsection{Selective literature review}

Sensor Network Localization (SNL) is a fundamental problem in signal processing.
{While a comprehensive review is beyond the scope of this paper, we briefly highlight the lines of work most relevant to this study.}
From a methodological perspective, most of the existing research can be classified into two categories: vector-based and matrix-based.
The vector-based category often makes use of  
coordinate-based optimization, including the constrained weighted least squares approach \citep{Re_CWLS_2006, Xu2025}, the sequential weighted least squares algorithm (SWLS) \citep{Re_SWLS_2008}, and the two-stage weighted least squares approach \citep{Re_2SWLS_2016}, among others \citep{larsson2025single, Geng2025}. 
In contrast, the matrix-based category makes use of semidefinite programming (SDP) \citep{Re_YeSDP_2004, Re_YeSDP_2006, Re_Wu_2023} and  Euclidean distance matrix (EDM) optimization \citep{Re_QiSNT_2013, Re_QiYuan_2014, bai2016tackling}. This class of methods has attracted increasing attention due to their capability of modeling
complex settings in terms of conic optimization \citep{li2017inexact, lu2020feasibility}. 
EDM optimization, in particular, has proven to be an efficient tool, as it can directly utilize observed Euclidean distances in the form of linear constraints \citep{Re_QiYuan_2014, Re_QiLag_2013} and hence it significantly reduces model complexity (e.g., avoiding quadratic constraints often appeared in the vector-based models).
Furthermore, \citet{Re_Zhoubox_2018} proposed a fast matrix majorization projection method, while  \citet{Re_Shi_2023} developed a facial reduction approach, both of which can efficiently tackle SNL problems by making full use of EDM-based formulations.
{
Another important advantage of matrix-based optimization is its ability to exploit the underlying low-rank structure, which has been widely studied over the past decade 
\citep{cambier2016robust, zhang2019localization, wang2025robust}. In particular, robust matrix-based variants, including formulations in $\ell_p$ spaces with $p<2$ and models based on robust entry-wise loss functions, have been shown to mitigate the influence of outliers and heavy-tailed noise in matrix completion and large-scale network localization problems  
 \citep{zeng2017outlier, li2022fast}. These works are closely related to the present study in that they also exploit matrix-based geometric structure. Compared to the existing work, this paper focuses on incorporating informative 3D angle constraints into an EDM framework through explicit distance bounds.
}

When it comes to SSLAR, the above-mentioned approaches are capable of handling range constraints but face challenges in dealing with the angle information. 
For example, there seems to be no trivial extension of SWLS to solve SSLAR. 
Existing research with angle information mainly focuses on the vector-based formulation. 
For instance, Aubry et al. introduced an angle constrained least squares method in two-dimensional (2D) \citep{Re_2dAubry_2020} and developed an angle and range constrained estimator (ARCE) in three-dimensional (3D) \citep{Re_Enhance_a}. 
\citet{Re_Marino_2024} further investigated combining ARCE with a scalable sum-product algorithm \citep{Re_Brambilla} to accomplish localization and multi-target tracking tasks. 
For matrix-based formulation, \citet{Re_2dBiswas_2005} proposed an SDP-based algorithm 
that uses angle information, specifically targeting 2D scenarios and leveraging the cosine law. 
For moving-target localization,  \citet{Re_Jia_2025} proposed a closed-form solution approaching the Cramer-Rao lower bound and a semidefinite relaxation for joint localization and calibration, 
also in a 2D context.  
These matrix-based methods are limited to 2D settings due to the difficulties in capturing angle constraints in higher dimensions.
{
Moreover, many representative low-rank and matrix completion based localization methods are developed for settings with only range measurements \citep{zhang2019localization}, and therefore are not directly comparable with the present SSLAR setting. Accordingly, the main numerical comparisons in this paper focus on methods that use the same angle and range information, so as to ensure a fair comparison under the same prior information.}
The main purpose of this paper is to resolve this challenging task that simultaneously incorporates both the angle and the range information in an EDM optimization model in 3D SSL with an efficient algorithm.

\subsection{Main contributions} 
In this paper, we propose a robust EDM optimization framework for 3D SSLAR. 
The main contributions are summarized as follows:
\begin{itemize}
    \item {We introduce EDM optimization to the 3D SSLAR, which incorporates range measurements and angle information into a unified distance-based matrix optimization model.} 
    In this formulation, the existing 3D angle information is incorporated into EDM optimization through explicit lower and upper distance bounds. 
    This is achieved by solving a set of 2D constrained optimization subproblems, which have a finite number of KKT (Karush-Kuhn-Tucker) points. 
    The resulting bounds bridge the gap between angle-based measurements and distance-based EDM models.

    \item {The proposed EDM formulation permits robust localization under noisy measurements in terms of the least absolute deviation modeled by the $\ell_1$-norm.} 
    Unlike vector-based approaches in \citep{Re_Enhance_a}, where the $\ell_1$-norm leads to nonsmooth and nonconvex objectives with undesired computational complexity, our EDM formulation 
    {naturally supports the robust $\ell_1$-norm criterion} 
    within the matrix optimization framework. 
    We note that \citep{Re_Enhance_a} used the squared $\ell_2$-norm for its least-squares formulation, leading to a smooth objective. 
    We also consider EDM optimization with the least-squares formulation, which is computationally less challenging than the robust $\ell_1$-norm formulation.

    \item Based on the above formulation, we develop a practical algorithm framework, including a multi-start initialization strategy and a majorization penalty solver. 
    Numerical comparisons confirm that our framework outperforms state-of-the-art vector-based solvers in localization accuracy while maintaining competitive computational efficiency, particularly in low Signal-to-Noise Ratio (SNR) scenarios.
\end{itemize}

\subsection{Organization}

The paper is organized as follows. 
In Section~\ref{Section-Preliminary}, we review the vector-based model of SSLAR in 3D
and introduce the box-constrained EDM optimization (matrix-based model), highlighting its principle, advantages and challenges in using the model. 
{
Section~\ref{Section-Angle} presents how the existing angle information is incorporated into the EDM model through lower and upper distance bounds that form box constraints on the source-to-sensor distances.} 
In Section~\ref{Section-Algorithm}, we show how the majorization penalty method developed in
\citep{Re_Zhoubox_2018, Re_Zhourobust_2018} can be adapted to our EDM optimization problem.
Section~\ref{Section-Numerical} focuses on numerical implementation and comparison.
Final conclusions are in Section~\ref{Section-Conclusion}.


\textbf{Notation:}
We let $\mathbb{R}^n$ denote the $n$-dimensional Euclidean space endowed with the standard inner product $\langle \cdot,\cdot\rangle$. The induced norm is the Euclidean norm $\|\cdot\|$, also known as the $\ell_2$-norm. {Bold lowercase letters, such as $\mathbf{v}\in\mathbb{R}^n$, denote column vectors, and bold uppercase letters, such as $\mathbf{D}$, denote matrices.} The $\ell_1$-norm of a vector $\mathbf{v}$ is defined by $\|\mathbf{v}\|_1:=|v_1|+\cdots+|v_n|$. Let $\mathcal{S}^n$ denote the space of $n\times n$ real symmetric matrices endowed with the standard trace inner product and the induced Frobenius norm, again denoted by $\|\cdot\|$. The set of all positive semidefinite matrices in $\mathcal{S}^n$ is denoted by $\mathcal{S}_+^n$. {We further define $\mathbf{1}:=(1,\ldots,1)^{\top}$ and $\mathbf{1}^{\perp}:=\{\mathbf{v}\in\mathbb{R}^n:\mathbf{1}^{\top}\mathbf{v}=0\}$, as well as the conditionally positive semidefinite cone $\mathcal{K}_+^n:=\{\mathbf{A}\in\mathcal{S}^n:\mathbf{v}^{\top}\mathbf{A}\mathbf{v}\ge 0,\ \forall\,\mathbf{v}\in\mathbf{1}^{\perp}\}$. For two matrices $\mathbf{A}$ and $\mathbf{B}$ of the same size, $\mathbf{A}\circ\mathbf{B}$ denotes their Hadamard (entrywise) product, i.e., $(\mathbf{A}\circ\mathbf{B})_{ij}:=A_{ij}B_{ij}$.
We use $\diag(\mathbf{A})$ to denote the vector of diagonal entries of a square matrix $\mathbf{A}$, and $\Diag(a_1,\ldots,a_n)$ to denote the corresponding diagonal matrix.
For a nonnegative matrix $\mathbf{A}$, $\sqrt{\mathbf{A}}$ denotes the matrix obtained by taking the square root entrywise, i.e., $(\sqrt{\mathbf{A}})_{ij}:=\sqrt{A_{ij}}$. For a matrix $\mathbf{A}$, $\|\mathbf{A}\|_1$ denotes the entrywise $\ell_1$ norm, i.e., $\|\mathbf{A}\|_1:=\sum_{i,j}|A_{ij}|$.}
Finally, the notation ``$:=$'' means ``is defined as''.

\section{Vector model and EDM model} \label{Section-Preliminary} 

In this section, we review the vector-based model for the 3D SSLAR in \citep{Re_Enhance_a} and propose an EDM model. We will emphasize their differences.

\subsection{The vector model}

This part is taken from \citep{Re_Enhance_a} with details omitted.
A radar system consisting of $m$ nodes (one transmitter and $(m-1)$ receivers) is 
set up to estimate 
the position of an unknown target, as illustrated in Fig.~\ref{fig_intro}. 
The coordinate system is built as follows. 
Without loss of generality, the transmitter (the active radar) $\bfx_1$ is placed at the origin, 
i.e., $\mathbf{x}_1 = (0, 0, 0)^{\top} \in \mathbb{R}^3$.
The $i$-th receiver is positioned at $\mathbf{x}_i = (x_i, y_i, z_i)^{\top} \in \mathbb{R}^3$ for $i = 2, \ldots, m$. 
The unknown target  is denoted by $\mathbf{x} = (x, y, z)^{\top} \in \mathbb{R}^3$. 
There are a total of $n:=(m+1)$ points.

\begin{figure}[t]
	\centering
	\includegraphics[width=0.9\linewidth]{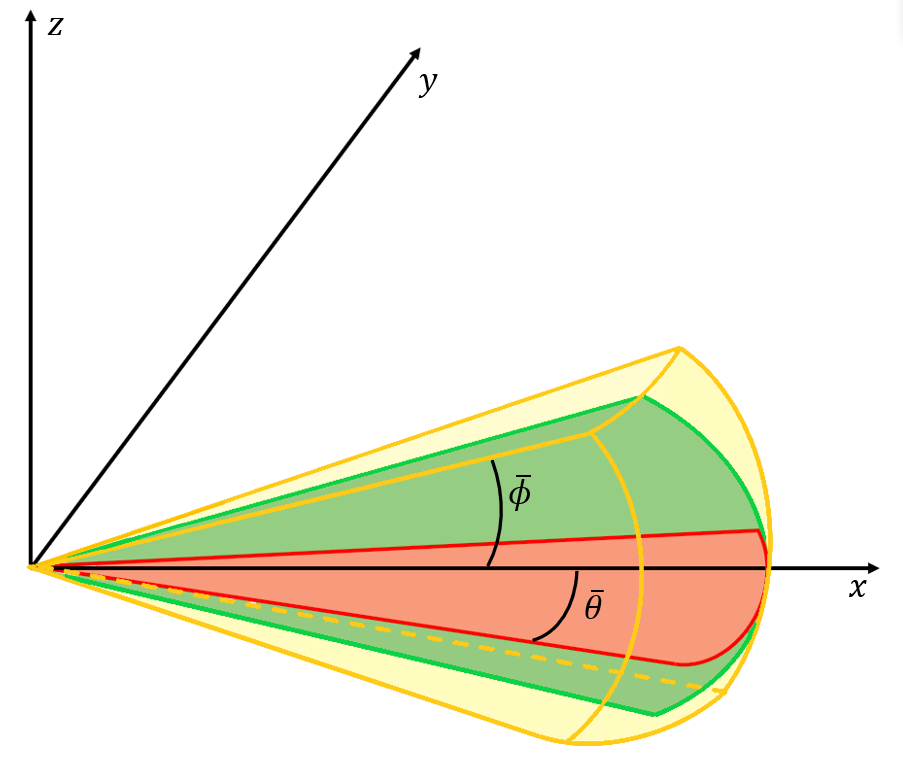}
	\caption{Representation of the antenna beamwidth (adapted from \citep{Re_Enhance_a}).}
	\label{fig_beam}
\end{figure}

To perform the measurement process, the transmitter employs an antenna characterized by a directional transmit/receive beam pattern with a given main-lobe width and range. 
The beam is assumed to be steered along the $x$-axis of the reference coordinate system. 
Let us denote the lower and upper bounds of the detectable range bin by $r_L$ and $r_U$, respectively. 
Denote the antenna beamwidths in the $(x, y)$ and $(x, z)$ planes by $\bar{\theta}$ and $\bar{\phi}$, respectively, as illustrated in Fig.~\ref{fig_beam}.

The azimuth and elevation angles of the target are defined as\footnotemark
\footnotetext{
	$
 \operatorname{atan2}(y, x) =
\begin{cases}
  \arctan\left(\frac{y}{x}\right), & \text{if } x > 0, \\
  \arctan\left(\frac{y}{x}\right) + \pi, & \text{if } x < 0 \text{ and } y \ge 0, \\
  \arctan\left(\frac{y}{x}\right) - \pi, & \text{if } x < 0 \text{ and } y < 0, \\
  \frac{\pi}{2}, & \text{if } x = 0 \text{ and } y > 0, \\
  -\frac{\pi}{2}, & \text{if } x = 0 \text{ and } y < 0, \\
  \text{not defined}, & \text{if } x = 0 \text{ and } y = 0.
\end{cases}
	$}
\begin{equation} \label{eq_twoangle}
	\theta := \operatorname{atan2}(y, x), 
	\quad \phi := \operatorname{atan2}(z, x).
\end{equation}
For a target $\mathbf{x}$ to be illuminated by the transmitter, it must satisfy the range constraint given by 
$
r_L \leq \|\mathbf{x}\| \leq r_U,
$
and the angle constraints given by
\begin{equation}\label{angle1}
	-\bar{\theta} \leq \theta \leq \bar{\theta}, \quad -\bar{\phi} \leq \phi \leq \bar{\phi},
\end{equation}
{ where $\bar{\theta}$  and $\bar{\phi}$ are the half-side antenna beamwidths and satisfy the physical constraints
$0 \le \bar{\theta} < \frac{\pi}{2}$ and $0 \le \bar{\phi} < \frac{\pi}{2}$.}


Equation \eqref{angle1} can be equivalently rewritten as 
\begin{equation}\label{angle2}
	x > 0, \
	-\gamma_{a} x \leq y \leq \gamma_{a} x, \
	- \gamma_{e} x \leq z \leq \gamma_{e} x,
\end{equation}
where $\gamma_{a} := \tan \bar{\theta}$ and $\gamma_{e} := \tan \bar{\phi}$.

When the target resides within this coverage, the radar nodes acquire noisy range measurements $\delta_{in}$, which are assumed to be given by
(recall $\bfx$ is the unknown source and $n=m+1$)
\[
\delta_{in}  = \|\mathbf{x}_{i} - \mathbf{x}\| + \epsilon_{i}, \ i= 1, \ldots, m,
\]
where $\epsilon_{i}$ represents measurement noise.  
For the details how $\delta_{in}$ were actually measured, see \citep[Part~II]{Re_Enhance_a}.

The (least-squares) vector model \citep[Eq.~(12)]{Re_Enhance_a} can be equivalently stated as follows:
\begin{equation}\label{LSformulation}
	\begin{array}{rl}
		\min\limits_{\mathbf{x}\in \mathbb{R}^3} & f_2(\bfx) := \sum_{i=1}^m ( \| \mathbf{x} - \mathbf{x}_{i} \|^2 - \delta_{in}^2 )^2  \\
		\text{s.t.} & \|\mathbf{x}\|^2 = b^2, \qquad\qquad ~~~ \text{(range constraint)} \\
		& - \gamma_a x \leq y \leq  \gamma_a x,  \qquad \text{(azimuth constraint)} \\
		& - \gamma_e x \leq z \leq \gamma_e x,   \qquad \text{(elevation constraint)} \\
		& x > 0, \qquad \qquad \quad ~~~~ \text{(orientation constraint)}
	\end{array}
\end{equation}	
where $b := \max \big( \min(\delta_{1n}, r_{U}), r_{L} \big)$.
The vector model deserves some comments as follows.

\begin{remark} \label{Remark-Vector-Model}
	{\bf (i) On the range constraint.} We note that $\delta_{1n} \approx \| \bfx \|$ due to placing the transmitter at the origin, because the transmitter is powerful enough that its measurement $\delta_{1n}$ contains only a low level of noise. When $\delta_{1n}$ is projected to the range bin $[r_L, r_U]$, the noise would be removed resulting $\| \bfx\| = b$ as the range constraint, see the first paragraph of \citep[Part~III]{Re_Enhance_a} for more explanation.
	{\bf (ii) On the objective function.} The objective is the squared $\ell_2$-norm of the vector of squared differences between $\| \bfx - \bfx_i\|^2$ (the true squared distance) and $\delta_{in}^2$ (the squared measurement). This ``double'' squared-metric favors large distances (e.g., large distances often contain large noises). A more robust metric is the $\ell_1$-norm based:
    \begin{equation}\label{eq:l1norm}
	f_1(\bfx) := \sum\limits_{i=1}^{m} \Big| \| \bfx - \bfx_{i} \| - \delta_{in} \Big|.
    \end{equation}
	However, replacing $f_2$ by $f_1$ in \eqref{LSformulation} would make the vector model very
	difficult to solve because $f_1(\bfx)$ is nonsmooth and nonconvex, coupled with the nonconvex range
	constraint. We will see that the EDM model can handle the $\ell_1$-norm without causing too much numerical difficulty.	
\end{remark}		

\subsection{EDM model}

We note that the observed range information $\delta_{in}$ is measured in the Euclidean norm.
The basic theory of Euclidean geometry says that if the exact Euclidean distances between a set of points are given, then their positions are uniquely determined subject to elementary operations such as shifting, reflection and rotation (i.e., orthogonal transformations). This rigidity theory has been widely
used in sensor network localization \citep{so2007theory}.
Since our radar system has already been placed, the final position of the unknown source can be obtained through the
Procrustes procedure 
\citep[Chapter~20]{borg2005modern} and \citep{Re_QiLag_2013}.
Central to this rigidity theory is the concept of EDM \citep{dokmanic2015euclidean}. 
We introduce EDM in the context of SSLAR.

\subsubsection{When the range measurements are noise-free}

Suppose the range measurements $\delta_{in}$ are accurate for $i=1, \ldots, m$. 
Since the locations $\mathbf{x}_i$ of all $m$ receivers are known, we construct the following matrix consisting of the squared Euclidean distances among those points:
\be \label{Matrix-D}
\begin{array}{l}
	\mathbf D=
	\left[\begin{array}{cccc}
		0 & \cdots & \|\mathbf{x}_1 - \mathbf{x}_{m}\|^2 &  \delta_{1n}^2\\
		\|\mathbf{x}_{2} - \mathbf{x}_1\|^2 & \cdots & \|\mathbf{x}_{2}-\mathbf{x}_{m}\|^2 & 
		\delta_{2n}^2\\
		\vdots & \ddots &\vdots &\vdots \\
		\|\mathbf{x}_{m} - \mathbf{x}_1\|^2  &\cdots & 0 &  \delta_{mn}^2\\
		\delta_{1n}^2 &\cdots & \delta_{mn}^2 & 0
	\end{array}\right].
\end{array}
\ee
Since $\delta_{in}$, $i=1,\ldots,m$, are accurate, the matrix $\mathbf D$ is an EDM and satisfies the following two properties \citep{Re_Gower_1985, Re_QiLag_2013}:
\be\label{EDM-Conditions}
\diag(\mathbf D) = 0 \quad \mbox{and} \quad  -\mathbf D \in \mathcal{K}^n_+,
\ee
where $\mathcal{K}^n_+$ denotes the conditionally positive semidefinite cone defined in the Notation section.
Consequently, the matrix $\mathbf H := -\mathbf J \mathbf D \mathbf J/2$ is positive semidefinite, where $\mathbf J := \mathbf I -\frac{1}{ n} \mathbf 1 \mathbf 1^{\top}$ and $\mathbf I$ is the identity matrix of size $n$. 
The following eigenvalue-eigenvector decomposition is well defined (recall $n=m+1$):
\[
\mathbf H = \mathbf P \mathbf \Lambda \mathbf P^{\top}, \quad [\mathbf y_1, \ldots, \mathbf y_m, \mathbf y_n] := \mathbf \Lambda^{\frac{1}{2}} \mathbf P^{\top},
\]
where $\mathbf \Lambda := \Diag(\lambda_{1}, \ldots, \lambda_{r})$ with $\lambda_{1} \geq \ldots \geq \lambda_{r} > 0$ being the positive eigenvalues of $\mathbf H$, and $\mathbf P \in \mathbb{R}^{n \times r}$ contains the corresponding eigenvectors.
The basic theory of EDM (see, e.g., \citep{Re_Gower_1985}) ensures that, in our case, $r=3$ (the rank of $\mathbf H$), and the embedding points $\mathbf y_1, \ldots, \mathbf y_m$ can be mapped to the existing positions $\mathbf x_1, \ldots, \mathbf x_m$
through a mapping $\mathcal T$. A general formula for $\mathcal T$ can be found in \citep[Prop.~4.1]{Re_QiLag_2013}.
Therefore, $\mathcal T(\mathbf y_n)$ recovers the true position of the unknown source $\mathbf x$.

\subsubsection{When the range measurements are noisy}\label{sec-noisy}
This is the situation we will mainly deal with in this paper.
Since the measurements of $\delta_{in}$ are contaminated with noises, the matrix
$\bf D$ in \eqref{Matrix-D} is not EDM anymore. To signify this, we denote it by the matrix
$\mathbf \Delta$. A natural idea was to compute an EDM $\bf D$ that is closest to $\mathbf \Delta$, resulting in
the nearest EDM problem studied in \citep{Re_QiSNT_2013}. 
However, we have more information, such as angles formed by beams, to build into such a problem. 
We explain how we achieve this. 

Let $\bf D$ denote a true EDM. We would like to seek one such $\bf D$ that is closest to $\mathbf \Delta$ satisfying
certain properties. 
Firstly, we define the closeness. 
We choose $\ell_1$-norm to measure it due to its robustness in embedding \citep{Re_Zhourobust_2018} (recall $f_1(\bfx)$ in \eqref{eq:l1norm}):
\begin{align*}
	F_1(\mathbf{D}) &:=  \sum_{i=1}^m | \| \bfx - \bfx_i\| - \delta_{in} | \\
	&=\sum_{i=1}^m | \sqrt{D_{in}} - \delta_{in} |
	= \frac{1}{2}\| \sqrt{\mathbf{D}} - \sqrt{\mathbf{\Delta}} \|_1,
\end{align*}
where we fixed the top $m\times m$ block of $\bf D$ (i.e., $D_{ij} = \Delta_{ij} = \| \bfx_i - \bfx_j \|^2$ for all $i, j=1, \ldots, m$). Those measurements are already available and accurate.
Obviously, we can also adopt the (squared) least-square objective in $f_2(\bfx)$ in \eqref{LSformulation}:
\[
F_2(\mathbf{D}) := \sum_{i=1}^m (D_{in} - \delta_{in}^2 )^2 = \frac{1}{2}\| \mathbf{D} -\mathbf{\Delta} \|^2 .
\]
This objective is convex and differentiable, but it favors large distances.

Secondly, we specify the conditions that $\bf D$ should satisfy. 
Obviously, it must be an EDM satisfying the two properties in \eqref{EDM-Conditions}.
Moreover, its embedding dimension must be $r=3$. In other words, 
the rank of the matrix $\mathbf{H}=-\mathbf{JDJ}/2$ must not be greater than $r$. 
There is a good way to capture those properties. Let
\[
\K^n_+(r) := \K^n_+ \cap \left\{
\mathbf{D} \in \S^n \ | \ \mbox{rank}(\mathbf{H}) \le r
\right\} .
\]
This is known as the $r$-cut of the conditional positive semidefinite cone.

Finally, we reach our EDM optimization model for SSLAR:
\begin{equation}\label{BoxEDM}
	\begin{aligned}
		\min\limits_{\mathbf{D} \in \S^n} & F_p (\mathbf{D}) && \mbox{($p=1$ or $2$)}\\
		\mbox{s.t.} & -\mathbf{D} \in \mathcal{K}_{+}^{n}(r), &&\mbox{($r$-cut constraint)}\\
		& D_{ij} = \| \bfx_i - \bfx_j \|^2,  1\le i, j\le m, &&\mbox{(fixed constraints)}\\
		&	D_{1n} = b^2, &&\mbox{(range constraint)}\\
		& l_i \leq D_{in} \leq u_i, 2 \le i \le m.  &&\mbox{(angle constraints)}
	\end{aligned}
\end{equation}

\begin{remark}\label{Remark-EDM-Model}
	(i) The variable in \eqref{BoxEDM} is $\bf D$. The constraints on $\bf D$ are linear except the $r$-cut constraint, which ensures that the embedding dimension is $r=3$. 
	In the case $p=2$, the objective $F_2(\mathbf{D})$ is strongly convex. This allows efficient computation 
	because we can handle the $r$-cut well (more on this later).
	For $p=1$, the numerical procedure for $p=2$ can be modified to solve this robust case. 
	(ii) The fixed constraints and the range constraint have to be obeyed by $\bf D$ as those measurements are already available. 
	A strong claim here is that the angle measurements in the vector model \eqref{LSformulation} 
	can be represented as a box constraint: $D_{in} \in [l_i, u_i]$ with $0< l_i \le u_i$.
	This is the major task we will complete in the next section. 
	With this representation, the EDM model \eqref{BoxEDM} is well structured and will
	yield high quality localization. 
    {(iii) The feasible set of \eqref{LSformulation} is nonempty; for example,
$
\frac{b}{\sqrt{1+\gamma_a^2+\gamma_e^2}}(1,\gamma_a,\gamma_e)^\top
$
is a feasible point. Hence, \eqref{LSformulation} admits at least one global minimizer, since its feasible set is closed and bounded and its objective function is continuous. When problem \eqref{LSformulation} is reformulated in terms of Euclidean distance matrices, every feasible point of \eqref{LSformulation}, together with the sensors $\mathbf{x}_1,\ldots,\mathbf{x}_m$, generates a corresponding EDM. Therefore, the feasible set of \eqref{BoxEDM} is also nonempty, although the objective function of \eqref{BoxEDM} may differ from that of \eqref{LSformulation}.}
	(iv) Once we get the optimal $\bf D$, we can use the procedure stated in the previous section to get the
	final localization through the mapping $\T(\cdot)$.
\end{remark}

\section{Angle measurements as box constraints} \label{Section-Angle} 

This is the main section that derives the lower bound $l_i$ and upper bound $u_i$ in \eqref{BoxEDM}, $i = 2, \ldots, m$. 
Since these bounds encode critical angle information, it is of great importance to obtain $l_i$ and $u_i$ accurately and efficiently.

To this end, define
{
$$
\begin{aligned}
	\Omega := &\{\mathbf{x} = (x, y, z) ^{\top} \in \mathbb{R}^3 \mid \|\mathbf{x}\|^2 = b^2,\\
	& \ - \gamma_{a} x \leq y \leq  \gamma_{a}  x,\ - \gamma_{e}x \leq z \leq  \gamma_{e}x,\ x > 0\}.
\end{aligned}
$$}

We consider the following pair of subproblems:
\begin{equation}\label{eq_PL}
	\begin{array}{rl}
		l_i :=	\underset{\mathbf{x} \in \Omega}{\min }&\| \mathbf{x} - \mathbf{x}_{i} \|^2 
	\end{array}
	\tag{$\underline{\mathcal{P}}_{i}$}
\end{equation}
and 
\begin{equation}\label{eq_PU}
	\begin{array}{rl}
		u_i :=	\underset{\mathbf{x} \in \Omega}{\max}&\| \mathbf{x} - \mathbf{x}_{i} \|^2 .
	\end{array}
	\tag{$\overline{\mathcal{P}}_{i}$}
\end{equation}

Due to the non-convexity of the feasible set, standard optimization methods may only yield locally optimal solutions when directly solving subproblems \eqref{eq_PL} and \eqref{eq_PU}, which compromises the accuracy of the bounds. Moreover, repeatedly invoking solvers for each $i$ is computationally expensive. 

Fortunately, there is { a convenient} way to represent those problems in two dimensions through 
variable transformation. The resulting problems are much easier to handle.
Let
$\mathbf{v} := \left(v_1, v_2\right) ^{\top}$
with $v_{1} = \tan\theta$, $v_{2}= \tan\phi$. 
We have the following technical result, whose proof is in 
{\ref{Appendix-Lemma}}.

\begin{lemma}
	\label{lem:extremum}
	For $\mathbf{x} \in \Omega$, $\left\|\mathbf{x}_{i} - \mathbf{x}\right\|^{2}$ can be written as a function of  $\mathbf{v}$, denoted as $h_{i}\left(\mathbf{v}\right)$. The following results hold
	\begin{equation}
		\label{eq:h_{i}}
		h_{i}\left(\mathbf{v}\right) = -2 \dfrac{b}{\sqrt{1 + v_{1}^{2} + v_{2}^{2}}} \left(x_{i} + y_{i} v_{1} + z_{i} v_{2}\right) + \delta_{1i}^2 + b^2,
	\end{equation}
	and the gradient of $h_{i}\left(\mathbf{v}\right)$ takes the following form
	\begin{equation}	\label{eq:partial_h_{i}}
		\nabla h_{i}(\mathbf{v}) 
		= 2 b
		\left[
		\begin{array}{c}
			\dfrac{v_{1} \left(x_{i} + y_{i} v_{1} + z_{i} v_{2}\right) - y_i \left(1 + v_1^2 + v_2^2\right)}
			{\left(1 + v_{1}^{2} + v_{2}^{2}\right)^{\frac{3}{2}}}  \\
			\dfrac{v_{2} \left(x_{i} + y_{i} v_{1} + z_{i} v_{2}\right) - z_i \left(1 + v_1^2 + v_2^2\right)}
			{\left(1 + v_{1}^{2} + v_{2}^{2}\right)^{\frac{3}{2}}}
		\end{array}
		\right].
	\end{equation} 
\end{lemma}

Based on the variable transformation and Lemma \ref{lem:extremum}, the original problems $(\underline{\mathcal{P}}_{i})$ and $(\overline{\mathcal{P}}_{i})$ can be reformulated as 2D box-constrained smooth problems,
\begin{equation}\label{eq_minf}
	\begin{array}{rl}
		l_i=	\underset{\mathbf{v} \in V}{\min }&h_{i}\left(\mathbf{v}\right)
	\end{array}
	\tag{$\underline{\mathcal{P}}_{i}^{\prime}$}
\end{equation}
and
\begin{equation}\label{eq_maxf}
	\begin{array}{rl}
		u_i =	\underset{\mathbf{v} \in V}{\max }&h_{i}\left(\mathbf{v}\right),
	\end{array}
	\tag{$\overline{\mathcal{P}}_{i}^{\prime}$}
\end{equation}
where $V := \{ \mathbf{v} = \left(v_1, v_2\right)^{\top} \in \mathbb{R}^2 \mid -\gamma_{a} \leq v_{1} \leq \gamma_{a},\ -\gamma_{e} \leq v_{2} \leq \gamma_{e}\}.$ 
{It is important to note that the feasible set $V$ is a box.}
Therefore, the optimal solutions of both problems can be characterized by their KKT conditions
\citep{Re_Numerical}. 
{Furthermore, the number of KKT candidate points is finite, with at most nine cases to be examined.}
This result is stated in the following theorem, whose proof is in 
{\ref{Appendix-Thm}}.

\begin{theorem}\label{th_candidateset}
	The candidate points that satisfy the KKT conditions for \eqref{eq_minf}  and  \eqref{eq_maxf} are given in Table \ref{ta_candidate_L}, where the condition columns (the leftmost column and the rightmost column) mean that the corresponding condition must be satisfied so that the candidate solution in the third column is the KKT solution.
	\begin{table*}[!t]
    \tabfontsize 
		\centering
		\renewcommand{\arraystretch}{1.9} 
		\caption{KKT candidate solutions and conditions for  \eqref{eq_minf} and  \eqref{eq_maxf}.} \label{ta_candidate_L}
		\begin{tabular}{c|c|c|c|c}
			\Xhline{1.5pt} 
			Condition for \eqref{eq_minf} & Case & Candidates & Case & Condition for \eqref{eq_maxf}\\
			\Xhline{1pt} 
			$ -\gamma_{a} \leq \dfrac{y_{i}}{x_{i}} \leq \gamma_{a}$,  & \multirow{2}{*}{1} & \multirow{2}{*}{$ \left( \dfrac{y_{i}}{x_{i}}, \dfrac{z_{i}}{x_{i}}\right)^{\top}$} & \multirow{2}{*}{1} & $ -\gamma_{a} \leq \dfrac{y_{i}}{x_{i}} \leq \gamma_{a}$,  \\
			$-\gamma_{e} \leq \dfrac{z_{i}}{x_{i}} \leq \gamma_{e}$ & & & &
			$ -\gamma_{e} \leq \dfrac{z_{i}}{x_{i}} \leq \gamma_{e}$\\
			\Xhline{1pt} 
			$ -\gamma_{e} \leq  \dfrac{\left(1 + \gamma_{a}^{2}\right) z_{i}}{x_{i} - \gamma_{a} y_{i}} \leq \gamma_{e}$,	 & \multirow{2}{*}{2.1} &  \multirow{2}{*}{$\left(-\gamma_{a}, \dfrac{\left(1 + \gamma_{a}^{2} \right) z_{i}}{x_{i} - \gamma_{a} y_{i}}\right) ^{\top}$} & \multirow{2}{*}{2.1} & $ -\gamma_{e} \leq  \dfrac{\left(1 + \gamma_{a}^{2}\right) z_{i}}{x_{i} - \gamma_{a} y_{i}} \leq \gamma_{e}$,  \\
			$\gamma_{a}x_i + y_i < 0$ & & & &
			$\gamma_{a}x_i + y_i > 0$\\
			\Xhline{0.5pt} 
			$-\gamma_{e} \leq  \dfrac{\left(1 + \gamma_{a}^{2}\right) z_{i}}{x_{i} + \gamma_{a} y_{i}} \leq \gamma_{e}$, & \multirow{2}{*}{2.2} &  \multirow{2}{*}{$\left(\gamma_{a}, \dfrac{\left(1 + \gamma_{a}^{2}\right) z_{i}}{x_{i} + \gamma_{a} y_{i}}\right) ^{\top}$}  &  \multirow{2}{*}{2.2} & $-\gamma_{e} \leq  \dfrac{\left(1 + \gamma_{a}^{2}\right) z_{i}}{x_{i} + \gamma_{a} y_{i}} \leq \gamma_{e}$,  \\
			$\gamma_{a}x_i - y_i < 0$ & & & &
			$\gamma_{a}x_i - y_i > 0$\\
			\Xhline{0.5pt} 
			$-\gamma_{a} \leq \dfrac{\left(1+\gamma_{e}^{2}\right) y_{i}}{x_{i} - \gamma_{e} z_{i}}  \leq \gamma_{a}$,  & \multirow{2}{*}{2.3} & 	\multirow{2}{*}{$\left(\dfrac{\left(1+\gamma_{e}^{2}\right) y_{i}}{x_{i} - \gamma_{e} z_{i}}, -\gamma_{e}\right) ^{\top}$}  & \multirow{2}{*}{2.3} &  $-\gamma_{a} \leq \dfrac{\left(1+\gamma_{e}^{2}\right) y_{i}}{x_{i} - \gamma_{e} z_{i}}  \leq \gamma_{a}$, \\
			$\gamma_{e}x_i + z_i < 0$ & & & &
			$\gamma_{e}x_i + z_i > 0$\\
			\Xhline{0.5pt} 
			$-\gamma_{a} \leq \dfrac{\left(1+\gamma_{e}^{2}\right) y_{i}}{x_{i} + \gamma_{e} z_{i}}  \leq \gamma_{a}$,  & \multirow{2}{*}{2.4}& \multirow{2}{*}{$\left(\dfrac{\left(1+\gamma_{e}^{2}\right) y_{i}}{x_{i} + \gamma_{e} z_{i}}, \gamma_{e}\right) ^{\top}$} &  \multirow{2}{*}{2.4}& $-\gamma_{a} \leq \dfrac{\left(1+\gamma_{e}^{2}\right) y_{i}}{x_{i} + \gamma_{e} z_{i}}  \leq \gamma_{a}$, \\
			$\gamma_{e}x_i - z_i < 0$ & & & &
			$\gamma_{e}x_i - z_i > 0$ \\
			\Xhline{1pt} 
			$\gamma_{a} x_{i} + y_{i} (1 + \gamma_{e}^2) - \gamma_{a} \gamma_{e} z_{i} < 0$,  & \multirow{2}{*}{3.1} &  \multirow{2}{*}{$\left(-\gamma_{a}, -\gamma_{e}\right) ^{\top}$}  &  \multirow{2}{*}{3.1} & $\gamma_{a} x_{i} + y_{i} (1 + \gamma_{e}^2) - \gamma_{a} \gamma_{e} z_{i} >0$, \\
			$\gamma_{e} x_{i} + z_{i} (1 + \gamma_{a}^2) - \gamma_{a} \gamma_{e} y_{i} < 0$ & & & & $\gamma_{e} x_{i} + z_{i} (1 + \gamma_{a}^2) - \gamma_{a} \gamma_{e} y_{i} > 0$\\
			\Xhline{0.5pt} 
			$\gamma_{a} x_{i} + y_{i} (1 + \gamma_{e}^2) + \gamma_{a} \gamma_{e} z_{i} < 0$,   &  \multirow{2}{*}{3.2} &   \multirow{2}{*}{$\left(-\gamma_{a}, \gamma_{e}\right) ^{\top}$}  &   \multirow{2}{*}{3.2} &  $\gamma_{a} x_{i} + y_{i} (1 + \gamma_{e}^2) + \gamma_{a} \gamma_{e} z_{i} > 0$,   \\
			$\gamma_{e} x_{i} - z_{i} (1 + \gamma_{a}^2) - \gamma_{a} \gamma_{e} y_{i} < 0$ & & & & $\gamma_{e} x_{i} - z_{i} (1 + \gamma_{a}^2) - \gamma_{a} \gamma_{e} y_{i} > 0$\\
			\Xhline{0.5pt} 
			$\gamma_{a} x_{i} - y_{i} (1 + \gamma_{e}^2) - \gamma_{a} \gamma_{e} z_{i} < 0$,  &  \multirow{2}{*}{3.3} &  \multirow{2}{*}{$\left(\gamma_{a}, -\gamma_{e}\right) ^{\top}$} &   \multirow{2}{*}{3.3} &  $\gamma_{a} x_{i} - y_{i} (1 + \gamma_{e}^2) - \gamma_{a} \gamma_{e} z_{i} > 0$,  \\
			$\gamma_{e} x_{i} + z_{i} (1 + \gamma_{a}^2) + \gamma_{a} \gamma_{e} y_{i} < 0$ & & & & $\gamma_{e} x_{i} + z_{i} (1 + \gamma_{a}^2) + \gamma_{a} \gamma_{e} y_{i} > 0$\\
			\Xhline{0.5pt} 
			$\gamma_{a} x_{i} - y_{i} (1 + \gamma_{e}^2) + \gamma_{a} \gamma_{e} z_{i} < 0$,  &  \multirow{2}{*}{3.4} &   \multirow{2}{*}{$ \left(\gamma_{a}, \gamma_{e}\right) ^{\top}$} &   \multirow{2}{*}{3.4} &  $\gamma_{a} x_{i} - y_{i} (1 + \gamma_{e}^2) + \gamma_{a} \gamma_{e} z_{i} >0$, \\
			$\gamma_{e} x_{i} - z_{i} (1 + \gamma_{a}^2) + \gamma_{a} \gamma_{e} y_{i} < 0$ & & & & $\gamma_{e} x_{i} - z_{i} (1 + \gamma_{a}^2) + \gamma_{a} \gamma_{e} y_{i} > 0$  \\
			\Xhline{1.5pt} 
		\end{tabular}
	\end{table*}	
   \
\end{theorem}

\begin{remark}
Although the problems \eqref{eq_minf} and \eqref{eq_maxf} are generally non-convex, {Theorem~\ref{th_candidateset} provides a finite and complete characterization of all KKT candidate points. Therefore, by evaluating the objective function at these candidate points, the global minimum of \eqref{eq_minf} and the global maximum of \eqref{eq_maxf} are guaranteed to be attained.}
\end{remark}
We end this section by summarizing the calculation of the lower and upper bounds in Algorithm \ref{alg:ComputeLU}.	


\begin{algorithm}[ht]
	\caption{Compute lower and upper bound  $\mathbf{l}$ and $\mathbf{u}$}
	\label{alg:ComputeLU}
	\renewcommand{\algorithmicrequire}{\textbf{Input:}}
	\renewcommand{\algorithmicensure}{\textbf{Output:}}
	\begin{algorithmic}[1]
		\REQUIRE  $\bar{\phi}, \bar{\theta}$, and  $h_{i}(\mathbf{v}), \ i = 2, \ldots, m.$
		\ENSURE $\mathbf{l}$, $\mathbf{u} \in \mathbb{R}^{m-1}$. 
		\FOR{each index $i$}
		\STATE \textbf{S1:}
		Calculate all feasible candidate points in Table \ref{ta_candidate_L}.
		\STATE \textbf{S2:}
		For each candidate point $\mathbf{v}$, compute $h_{i}(\mathbf{v})$.
		\STATE \textbf{S3:}
		$\displaystyle l_{i} \gets \min \{ h_{i}(\mathbf{v}) \mid \mathbf{v}$   satisfies 
		condition for \eqref{eq_minf} in Table \ref{ta_candidate_L}$\}$,
		$\displaystyle u_{i} \gets \max \{ h_{i}(\mathbf{v}) \mid \mathbf{v}  $ satisfies  condition for \eqref{eq_maxf} in Table \ref{ta_candidate_L}$\}$.
		\ENDFOR
		\STATE \textbf{return} $\mathbf{l}, \mathbf{u}.$
	\end{algorithmic}
\end{algorithm}

\section{The EDMAR algorithm}  \label{Section-Algorithm} 

Having derived the lower bound $\mathbf{l}$ and the upper bound $\mathbf{u}$,
problem \eqref{BoxEDM} is well defined. It can be put in the following form:
\be \label{EDM-Optimization}
\min_{\mathbf{D} \in \B} \; F_p(\mathbf{D}) \quad \mbox{s.t.} \ - \mathbf{D} \in \K^n_+(r),
\ee 
where $\B := [\mathbf{L}, \mathbf{U}]$ is the box constraint with $\mathbf{L}, \mathbf{U} \in \S^n$ defined by
(note $\mathbf{L}, \mathbf{U}$ are symmetric, we only define their upper parts and we note that
$n = m+1$)
\[
L_{ij} = U_{ij} = \left\{
\begin{array}{ll}
	\| \bfx_i - \bfx_j \|^2 & \mbox{for} \ 1 \le i \le j \le m\\
	b^2 & \mbox{for} \ i=1, j=n \\
	0    & \mbox{for} \ i=j=n 
\end{array} 
\right .
\]
and
\[
 L_{in} = l_i , \quad U_{in} = u_i \quad \mbox{for} \ i=2,\ldots, m.
\]
Problems of this nature have been extensively investigated in \citep{bai2016tackling, Re_Zhoubox_2018, Re_Zhourobust_2018}. To solve \eqref{EDM-Optimization}, we adopt the majorization penalty framework. We briefly explain it below.

\subsection{The penalty problem}

Instead of solving problem \eqref{EDM-Optimization} directly, we try to solve its penalty form.
We note the following fact \citep{Re_QiYuan_2014}, 
$$
-\mathbf{D} \in \mathcal{K}_{+}^{n}(r) \iff g(\mathbf{D}) := \frac{1}{2}\left\|\mathbf{D}+\Pi_{\mathcal{K}_{+}^{n}(r)}(-\mathbf{D})\right\|^{2} = 0,
$$
where $\Pi_{\K^n_+(r)}(\mathbf{Z})$ denotes an orthogonal projection of $\mathbf{Z}$ onto the set $\K^n_+(r)$.
Although the projection is not unique, the function $g(\mathbf{D})$ is well defined and is concave.
Therefore, problem \eqref{EDM-Optimization} can be equivalently rewritten as
\[
\underset{\mathbf{D} \in \mathcal{B}}\min \; F_{p}(\mathbf{D}) \quad 
\text {s.t.}\quad  g(\mathbf{D}) = 0 .
\]
This yields the following penalty problem: 
\begin{equation}\label{Penalty}
	\underset{\mathbf{D} \in \mathcal{B}} \min \   F_{p}(\mathbf{D})+ \rho g(\mathbf{D})
\end{equation}
where $\rho >0$ is the penalty parameter.
The task now is to solve this 
penalty problem. For the functions $F_1$ and $g$, both are nonconvex and nondifferentiable. We use the popular majorization-minimization (MM) \citep{Re_Sun_2017, Varshney2026}
 technique to handle them. 
The idea is simple. Suppose we have a hard function $f(\bfx)$ to minimize,
we may update the current iterate $\bfx^k$ through minimizing its majorization function
at $\bfx^k$:
\[
  \bfx^{k+1} \in \arg\min \tilde{f}(\bfx; \bfx^k)
\]
where the majorization function $\tilde{f}(\bfx; \bfx^k)$ satisfies the property
\[ 
  \tilde{f}(\bfx; \bfx^k) \ge f(\bfx), \ \forall \ \bfx \quad 
  \mbox{and} \quad
  \tilde{f}(\bfx; \bfx^k) = f(\bfx^k).
\] 
This property guarantees $f(\bfx^{k+1}) \le f(\bfx^k)$, leading to convergence of the generated sequence under some metrics. 
We now construct the majorization functions for $F_1(\mathbf{D})$ and $g(\mathbf{D})$.

\subsection{Subproblem via majorization} 

\subsubsection{Majorization for $g(\mathbf{D})$}
This has been handled in \citep{Re_Zhoubox_2018} by using the
concavity of $g(\cdot)$.
\[ 
g(\mathbf{D}) \le \underbrace{g(\mathbf{D}^k) + \left\langle \Pi_{\mathcal{K}_{+}^{n}(r)} (-\mathbf{D}^k) , \mathbf{D} -\mathbf{D}^k\right\rangle}_{:= \tilde{g}(\mathbf{D}; \mathbf{D}^k)},
\] 
where $\mathbf{D}^k$ is the current iterate and $\Pi_{\K^n_+(r)}$ is a subgradient of $g(\cdot)$ at
$\mathbf{D}^k$ \citep{Re_QiYuan_2014}. Furthermore, $\Pi_{\K^n_+(r)}(\cdot)$ can be easily calculated \citep{Re_Zhoubox_2018} and $\tilde{g}(\mathbf{D}; \mathbf{D}^k)$ is a majorization of $g$ at $\mathbf{D}^k$.

\subsubsection{Majorization of $F_1(\mathbf{D})$}
There are a few ways to handle the absolute value function.
The one below gives us the best numerical results. 
It is based on the concavity of the square root function 
$\sqrt{x}$ for $x \ge 0$. It always holds:
\[
\sqrt{x} \leq \sqrt{\tilde{x}
} +\frac{x-\tilde{x}}{2\sqrt{\tilde{x}}}, \quad \mbox{for} \ \tilde{x} >0 .
\] 
For a given $\varepsilon>0$, we obtain for $i=2, \ldots, m$,
\begin{align*}
&\quad 	| \sqrt{D_{in}} - \delta_{in} |  < \sqrt{(\sqrt{D_{in}} - \delta_{in})^2 + \varepsilon  }\\
& \le 	\sqrt{(\sqrt{D^k_{in}} - \delta_{in})^2 + \varepsilon  }\\
&\quad     + \frac{(\sqrt{D_{in}} - \delta_{in})^2 + \varepsilon - ((\sqrt{D^k_{in}} - \delta_{in})^2 + \varepsilon)   }{2 \sqrt{(\sqrt{D^k_{in}} - \delta_{in})^2 + \varepsilon  } }
\end{align*}
\begin{align*}
& = \frac{(\sqrt{D_{in}} - \delta_{in})^2 }{2 \sqrt{(\sqrt{D^k_{in}} - \delta_{in})^2 + \varepsilon  }} + C_k, \qquad \qquad \qquad
\end{align*}
where $C_k$ is a constant independent of $\bf D$.
We now introduce a standard stabilization to replace $F_1(\mathbf{D})$ by
$$\widehat{F}_1(\mathbf{D}) := \sum_{i=1}^m \sqrt{(\sqrt{\mathbf{D}_{in}}-\delta_{in})^2 + \varepsilon},\quad \varepsilon>0,$$ 
whose majorization function in matrix form is 
$$\widetilde{G}(\mathbf{D};\mathbf{D}^k): = \|\sqrt{\widehat{\mathbf{W}}^k}\circ (\sqrt{\mathbf{D}}-\sqrt{\mathbf{\Delta}})\|^2 + \widehat{C}_k , $$
where
$$
\widehat{W}_{ij}^k =\left\{
\begin{array}{ll}
    \frac{1}{2\sqrt{(\sqrt{D^k_{ij} }- \delta_{ij})^2+\varepsilon}}, & i=2,\ldots,m, j=n,\\
    0,&\text{otherwise.}  
\end{array}
\right.
$$
{Note that  for each fixed $\bf D$, one has
$\widehat{F}_1(\mathbf{D})\to F_1(\mathbf{D}),\ \text{as }\ \varepsilon\downarrow 0$.
Accordingly, the convergence analysis for $p=1$ is carried out with respect to $\widehat{F}_1(\mathbf{D})$.}

\subsubsection{The subproblem to be solved}
Combining the two upper bounds gives the following subproblem: 

\begin{equation}\label{eq_mpsubproblem}
	\mathbf{D}^{k+1} := \arg\min_{\mathbf{D}\in \mathcal{B}}  \  {M_p(\mathbf{D};\mathbf{D}^k)},
\end{equation}
where for $p=1,2$, the function is given by
\begin{align*}
{M_1(\mathbf{D};\mathbf{D}^k)} &:= \|\sqrt{\widehat{\mathbf{W}}^k}\circ (\sqrt{\mathbf{D}}-\sqrt{\mathbf{\Delta}})\|^2+ \rho \tilde{g}(\mathbf{D}; \mathbf{D}^k),
 \\ 
{M_2(\mathbf{D};\mathbf{D}^k)} &:= \frac{1}{2}\|\mathbf{D}-\mathbf{\Delta}\|^2 + \rho \tilde{g}(\mathbf{D}; \mathbf{D}^k) .
\end{align*} 
Constant terms independent of $\bf D$  are dropped in the subproblem as they do not affect minimizers.

\subsection{\EDMAR~ algorithm and its convergence}
\subsubsection{Closed-form solution}

This part is to explain that the subproblem has a closed-form solution. First of all,
we note that the leading $m \times m$ block of $\mathbf{D}^{k+1}$ {is determined by the known geometry of the sensor network}:
\[
D_{ij}^{k+1} = \Delta_{ij}, \ 1\le i<j \le m, \ \
\mbox{and} \ \
D_{1n}^{k+1} = b^2.
\]
We only need to calculate $D_{in}^{k+1}$ for $i=2, \ldots, m$. 
The good news is that we only need to solve $(m-1)$ one-dimensional optimization
problem, which has a closed-form solution. We detail the formula below.

	For $p=1$, $D^{k+1}_{in}\ (i = 2, \ldots, m)$ is updated as follows:
	\be \label{D1-Update}
	\left\{
	\begin{array}{ll}
		\mathbf{\Delta}^k&:=-\frac{\widehat{\mathbf{W}}^k}{\rho} - \Pi_{\mathcal{K}_{+}^{n}(r)} (-\mathbf{D}^k) \\ [1ex]
		D^{k+1}_{in}&=\Pi_{[l_i,u_i]}({\tt{dcroot}}[{\Delta}^k_{in},\frac{\widehat{W}_{in}^k (\sqrt{\mathbf{\Delta}})_{in}}{\rho}])
	\end{array} 
	\right.
    ,
	\ee 
	where {\tt{dcroot}} is the root-finding formula used in \citep{Re_Zhoubox_2018}.

For $p=2$, $D^{k+1}_{in}\ (i = 2, \ldots, m)$ is updated as follows:
\be \label{D2-Update}
\left\{
\begin{array}{ll}
	\mathbf{\Delta}^k & := \frac{1}{\rho +1}(\mathbf{\Delta} -\rho \Pi_{\mathcal{K}_{+}^{n}(r)} (-\mathbf{D}^k)) \\ [1ex]
	D^{k+1}_{in} &= \Pi_{[l_i,u_i]}(\Delta^k_{in})
\end{array} 
\right.
.
\ee 
The algorithm is summarized in Algorithm~\ref{alg:MPAA}.

\begin{algorithm}
	\caption{$\text{EDMAR}_p$: EDM optimization with Angle and Range measurements}\label{alg:MPAA}
	\renewcommand{\algorithmicrequire}{\textbf{Input:}}
	\renewcommand{\algorithmicensure}{\textbf{Output:}}
	\begin{algorithmic}[1]
		\REQUIRE $\mathbf{x}_i,\ \delta_{in}, i =1,\ \ldots, m,\ \mathbf{\Delta},\ \mathbf{l},\ \mathbf{u}, \ \varepsilon,\ \rho>0$
		\ENSURE $\hat{\mathbf{x}}$
		\STATE \textbf{S1:} Initialize $\mathbf{D}^{0}$ and set $k:=0$.
		\STATE \textbf{S2:} Update $\mathbf{D}^{k+1} $ by \eqref{D1-Update} or \eqref{D2-Update}.	
		\STATE \textbf{S3:} Set $k \leftarrow k+1$ and repeat S2 until convergence, obtaining $\widehat{\mathbf{D}}$.
		\STATE \textbf{S4:} Apply the Procrustes process to $\widehat{\mathbf{D}}$ to obtain the final estimate $\hat{\mathbf{x}}$.
	\end{algorithmic}
\end{algorithm}

{
\subsubsection{Convergence analysis}
First, let $\Phi_p(\mathbf{D})$ denote the objective function of the penalty problem. Specifically, for $p=1$, we define
$\Phi_1(\mathbf{D}):=\widehat{F}_1(\mathbf{D})+\rho g(\mathbf{D}),$
whereas for $p=2$,
$\Phi_2(\mathbf{D}):=F_2(\mathbf{D})+\rho g(\mathbf{D}).$
Because the box constraint set
$
\mathcal{B}:=\{\mathbf{D}\in\mathcal{S}^n:\ \mathbf{L}\le \mathbf{D}\le \mathbf{U}\}
$
constructed by Algorithm~\ref{alg:ComputeLU} is nonempty, closed, and bounded, the penalty problem is well-defined on $\mathcal{B}$ and admits at least one optimal solution.

Next, we introduce the following definitions for our convergence results.

\begin{definition}[Stationary point of the penalty problem]
A matrix $\widehat{\mathbf{D}}\in\mathcal{B}$ is called a stationary point of the penalty problem if
\[
\left\langle \Gamma,\ \mathbf{D}-\widehat{\mathbf{D}} \right\rangle \ge 0,
\quad \forall \ \mathbf{D}\in\mathcal{B},
\]
for some $\Gamma\in \partial \Phi_p(\widehat{\mathbf{D}})$, where $\partial \Phi_p(\widehat{\mathbf{D}})$ denotes the subdifferential of $\Phi_p$ at $\widehat{\mathbf{D}}$.
\end{definition}

\begin{definition}[$\epsilon$-critical point of the original problem]
For a given $\epsilon>0$, a point $\widehat{\mathbf{D}}\in\mathcal{B}$ is called an $\epsilon$-critical point of the original constrained problem if
$
g(\widehat{\mathbf{D}})\le \epsilon
$
and
$
\left\langle \Gamma+\widehat\beta\,\widehat{\mathbf{D}}+\widehat\beta\,\Pi_{\mathcal{K}_{+}^{n}(r)}(-\widehat{\mathbf{D}}),\ \mathbf{D}-\widehat{\mathbf{D}} \right\rangle \ge 0,
\  \forall \ \mathbf{D}\in\mathcal{B},
$
for some $\widehat\beta\in\mathbb{R}$ and some $\Gamma\in \partial F_p(\widehat{\mathbf{D}})$.
\end{definition}

Following the majorization penalty framework in \citep[Thm.~2]{Re_Zhourobust_2018}, for any initialization $\mathbf{D}^0\in\mathcal{B}$, the sequence $\{\mathbf{D}^k\}$ generated by \EDMAR$_p$\ enjoys the following convergence properties:

\begin{enumerate}[(i)]
\item \textbf{Monotonicity:} The penalty objective sequence $\{\Phi_p(\mathbf{D}^k)\}$ is monotonically nonincreasing and bounded from below.

\item \textbf{Stationarity:} Every accumulation point of the sequence $\{\mathbf{D}^k\}$ is a stationary point of the corresponding penalty problem.

\item \textbf{$\epsilon$-Criticality:} For a sufficiently large penalty parameter $\rho$, every accumulation point of $\{\mathbf{D}^k\}$ is an $\epsilon$-approximate critical point of the original constrained EDM problem \eqref{EDM-Optimization}.

\item \textbf{Sequence convergence:} If an accumulation point is isolated, then the whole sequence $\{\mathbf{D}^k\}$ converges to that point.
\end{enumerate}

These properties provide the theoretical foundation for the \EDMAR$_p$\ algorithm. In particular, the above convergence statements hold for any initial point, that is, no local-neighborhood assumption on the initialization is required. However, since the penalty problem is still nonconvex, the uniqueness of the stationary point cannot be guaranteed. 
Therefore, the generated sequence is only guaranteed to converge to a stationary point of the corresponding penalty problem, rather than to a global minimizer.

This observation also motivates the multi-start strategy introduced in Section~\ref{Section-Numerical}, whose purpose is to reduce the sensitivity to initialization and to improve the practical robustness of the proposed algorithm. For a more comprehensive mathematical treatment of the $\epsilon$-criticality result, we refer the reader to \citep{Re_Zhoubox_2018, Re_Zhourobust_2018}.}
\section{Numerical results} \label{Section-Numerical}

	In this part, we conduct extensive numerical tests to verify the efficiency of the proposed model \eqref{BoxEDM} and the algorithm EDMAR$_p$. All tests are conducted on a MacBook Air with an Apple M3 chip (16 GB unified memory, 512 GB SSD) running macOS (Version 15.1.1). Our code is implemented in MATLAB R2024b.

    \subsection{Experimental setup}
\begin{figure}[t]
	\centering
	\includegraphics[width=\linewidth]{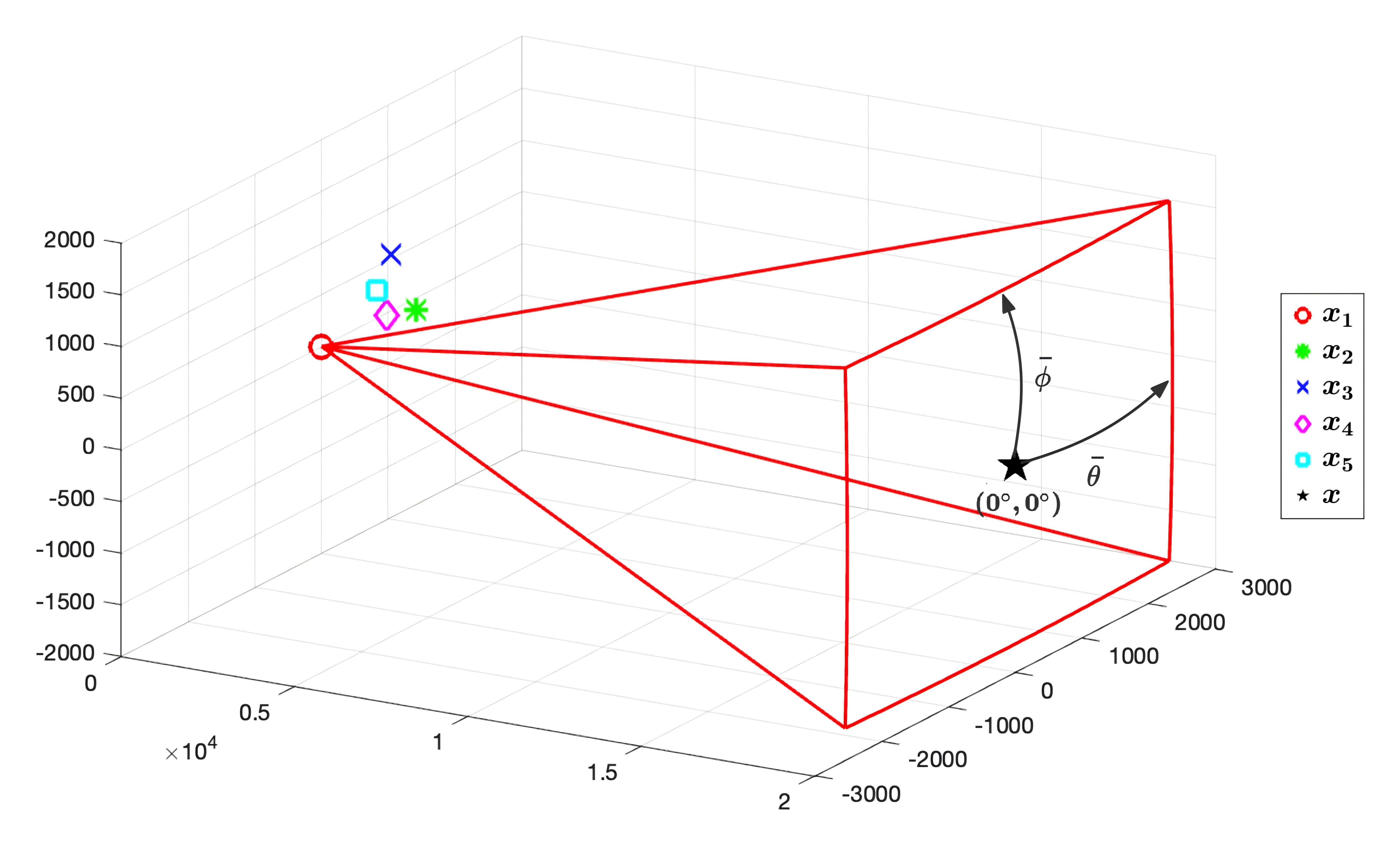}
	\caption{Geometric configuration of radar localization system in target location scenario, $(\theta,\phi) =(0^\circ,0^\circ)$.}
	\label{fig_3dmatlab}
\end{figure}

    \textbf{Test problem.}
{ 
Unless otherwise stated, we adopt the same 3D radar localization benchmark as in \citep{Re_Enhance_a} for a direct and fair comparison. Specifically, we use a network with $m=5$ nodes, including one transmitter located at $\mathbf{x}_1=(0,0,0)^{\top}$  $\mathrm{km}$
and four receivers located at
$\mathbf{x}_2=(0.916,0.941,0.095)^{\top}$  $\mathrm{km}$, $\mathbf{x}_3=(0.973,0.541,0.764)^{\top}$  $\mathrm{km}$, $\mathbf{x}_4=(0.955,0.483,0.191)^{\top}$  $\mathrm{km}$, and $\mathbf{x}_5=(0.936,0.350,0.477)^{\top}$  $\mathrm{km}$
    ($\mathrm{km}$ for kilometers).  
The remaining benchmark parameters are also chosen consistently with \citep{Re_Enhance_a}: the bandwidth is $B=2 \mathrm{~MHz}$, the loss factors are $ L_{1}=0\mathrm{~dB}$ and $L_{i}=6 \mathrm{~dB}$, $i=2, \ldots,m$, and the transmitter beamwidths are taken as either $(\overline{\theta}, \overline{\phi})= (7^\circ, 5^\circ)$ or $(\overline{\theta}, \overline{\phi})= (10^\circ, 7^\circ)$. The target locations are given by $$
    \mathbf{x}=(d\cos\theta \cos\phi, d\sin\theta \cos\phi, d\sin\phi) ^{\top},$$
    where $d = 20 \mathrm{~km}$, $(\theta,\phi) \in \{(0^\circ,0^\circ), (4^\circ,0^\circ), (6.9^\circ,4.9^\circ)\}$.
A representative configuration is shown in Fig.~\ref{fig_3dmatlab}, where the target positions are indicated by star markers.
For the experiments under different network sizes, the receiver locations are generated separately in the corresponding subsection.
}
 
     { In all experiments, the noisy range measurements $\delta_{in}$ in the EDM model are computed by}
    	$$\delta_{1n}  = \frac{1}{2}c \tau_1,\
    	\delta_{in}  = c \tau_i - \frac{1}{2} c \tau_1,\  i =2,\ldots,m,$$
    	where  $c$ is the speed of light, $\tau_i$ is the data of the time difference of arrival given in \citep[Part~II]{Re_Enhance_a}.

\textbf{Compared methods.}
\textit{Baseline strategies.} To evaluate the contribution of angle information, we propose two baseline strategies, EDMR$_1$ and EDMR$_2$, as simplified counterparts to EDMAR$_1$ and EDMAR$_2$. 
These methods retain the algorithmic framework of \EDMAR~ but do not use angle information. 
The box constraints $\mathbf{l}^{\prime} $ and $\mathbf{u}^{\prime} $ are defined as
$ l_i^{\prime}  = 0,  u_i^{\prime}  = \max_{1 \leq i,j \leq n} \Delta_{ij}, i = 2, \ldots, m$.
\textit{Existing benchmark methods.} 
 For comparison, we also include the state-of-the-art methods ARCE \citep{Re_Enhance_a} and the MATLAB solver fmincon (with the interior-point algorithm), both of which solve the problem \eqref{LSformulation} with $p=2$. All solver options are set to their stringent default values.
        
     \textbf{Initialization strategy.}
        { For the proposed \EDMAR$_p$ methods, we employ a deterministic multi-start initialization strategy to enhance the robustness and leverage the prior information $\mathbf{l}$ and $\mathbf{u}$ computed by Algorithm \ref{alg:ComputeLU}. This strategy is inspired by  \citep{Re_Marti_2018, Re_Marti_2019, Re_Hu_2020}.}  
			The set of initial points is constructed as follows
		$$\mathcal{E}^{n} = \left\{\mathbf{\Delta} \right\} \cup \left\{ \alpha \mathbf{L} + (1-\alpha) \mathbf{U} \mid \alpha \in \mathcal{A} \right\},$$
        where $\mathcal{A} \subset [0, 1]$ is a set of interpolation weights. In our experiments, let $\mathcal{A} = \{0,  0.5,  1\}$ and $\varepsilon = 0.1$.
        For each $\mathbf{D}^0 \in \mathcal{E}^n$, EDMAR$_p$ is executed with the penalty factor $\rho = n$ until the stopping conditions
	$$
	\operatorname{Fprog}_{k} \leq  5\sqrt{n}\times10^{-4} \text { and } \operatorname{Kprog }_{k} \leq 10^{-3}
	$$ are satisfied,
	where
    \begin{equation*}
\operatorname{Fprog}_{k}:=\frac{\Phi_{p}\left(\mathbf{D}^{k-1}\right)-\Phi_{p}\left(\mathbf{D}^{k}\right)}{1+\Phi_{p}\left(\mathbf{D}^{k-1}\right)}        
    \end{equation*}
    and
    \begin{equation*}
        \operatorname{Kprog}_{k}=1-\frac{\sum_{i=1}^{3}\left(\lambda_{i}^{2}-\left(\lambda_{i}-\max \left\{\lambda_{i}, 0\right\}\right)^{2}\right)}{\lambda_{1}^{2}+\ldots+\lambda_{n}^{2}},
    \end{equation*}
	with  $\lambda_{1} \geq \lambda_{2} \geq \ldots \geq \lambda_{n}$  are the eigenvalues of  $\left(-\mathbf{J} \mathbf{D}^{k} \mathbf{J}\right)$ .  

	We conduct the procedure through the mapping $\mathcal{T}(\cdot)$ to obtain the estimated position of the source, denoted by $\hat{\mathbf{x}}$. 
If $\hat{\mathbf{x}}$ satisfies the angle constraints~\eqref{angle2}, the process is terminated. Otherwise, the algorithm proceeds to the next initialization in $\mathcal{E}^n$. The initialization strategy is summarized in Algorithm~\ref{alg:InitStrategy}.

\begin{algorithm}
\caption{Multi-start Initialization Strategy}
\label{alg:InitStrategy}
\renewcommand{\algorithmicrequire}{\textbf{Input:}}
\renewcommand{\algorithmicensure}{\textbf{Output:}}
\begin{algorithmic}[1]
\REQUIRE $\mathbf{\Delta},\ \mathbf{L},\ \mathbf{U},\ \mathcal{A}$
\ENSURE $\hat{\mathbf{x}}$

\STATE Construct the initial point set:
\[
\mathcal{E}^n \leftarrow \left\{ \mathbf{\Delta} \right\} \cup \left\{ \alpha \mathbf{L} + (1 - \alpha) \mathbf{U} \mid \alpha \in \mathcal{A} \right\}
\]

\FOR{each $\mathbf{D}^0 \in \mathcal{E}^n$}
    \STATE Call Algorithm \ref{alg:MPAA}
    \IF{$\hat{\mathbf{x}}$ satisfies condition \eqref{angle2}}
        \RETURN $\hat{\mathbf{x}}$
    \ENDIF
\ENDFOR

\RETURN $\hat{\mathbf{x}}$ from the final iteration
\end{algorithmic}
\end{algorithm}

\textbf{Measuring the solution quality.}
For this purpose, we conduct $N=1000$ independent Monte Carlo simulations and adopt the following widely used measures: RMSE (Root Mean Square Error), Time, and Eigenratio.
RMSE is used to evaluate the estimation accuracy, defined as
\[
\mathrm{RMSE}=\sqrt{\frac{1}{N}\sum_{i=1}^{N}\|\hat{\mathbf{x}}_i-\mathbf{x}\|^2},
\]
where $\mathbf{x}$ is the true target position and $\hat{\mathbf{x}}_i$ is the estimated position in the $i$-th simulation.
The Time measure reports the average CPU time over $N$ independent trials. {For the proposed EDMAR$_p$ methods, this timing includes the full procedure of each trial, including the computation of the box constraints in Algorithm~\ref{alg:ComputeLU} and the iteration in Algorithm~\ref{alg:MPAA}.} Hence, it reflects the algorithm’s computational efficiency.
We define
\[
\text{Eigenratio}:=\frac{\sum_{i=1}^{3}|\lambda_i|}{\sum_{i=1}^{n}|\lambda_i|}.
\]
A ratio above $90\%$ indicates a high-quality EDM approximation.
{
\textbf{Computational complexity.}
The proposed framework consists of two main parts: the calculation of the box bounds in Algorithm~\ref{alg:ComputeLU} and the MM-based penalty iterations in \EDMAR$_p$. Recall that $m$ is the number of radar nodes and $n=m+1$.
Algorithm~\ref{alg:ComputeLU} processes the receivers independently. For each receiver, only a finite number of candidate points is generated and evaluated. Hence, the cost of computing the lower and upper bounds is $O(m)$.
The dominant cost of \EDMAR$_p$\ lies in the MM updates for the penalized EDM problem.
{
In the 3D setting, the embedding dimension is fixed at $r=3$, and each MM update is dominated by a truncated spectral decomposition associated with the projection onto $\mathcal{K}_+^n(r)$. Hence, the per-update complexity is $O(n^2)$. Under the fixed stopping tolerance and deterministic multi-start setting used in our experiments, the numbers of MM updates and initializations are regarded as constants independent of $n$. Therefore, the computational complexity of \EDMAR$_p$ is $O(n^2)$ with respect to the network size. 
For comparison, the ARCE algorithm in \citep{Re_Enhance_a} involves a root-search step and has computational complexity $O(n^2)$ for a fixed root-search accuracy.
}
}

{
To further clarify the differences between ARCE and the proposed EDMAR framework, 
Table~\ref{tab:arce-edmar} provides a systematic comparison in terms of formulation, robustness, solution guarantee, computational complexity, scalability, and possible extension.
}
\begin{table}[t]

\centering
\caption{Systematic comparison between ARCE and EDMAR.}
\label{tab:arce-edmar}
\small
\setlength{\tabcolsep}{4pt}
\renewcommand{\arraystretch}{1.15}
\begin{tabular}{p{0.35\linewidth}p{0.28\linewidth}p{0.28\linewidth}}
\hline
 & ARCE \citep{Re_Enhance_a} & EDMAR \\
\hline
Problem formulation 
& Eq.~\eqref{LSformulation} 
& Eq.~\eqref{BoxEDM} \\

Decision variable 
& $\mathbf{x}\in\mathbb{R}^3$ 
& $\mathbf{D}\in\mathcal{S}^n$ \\

Angle information
& \checkmark 
& \checkmark \\

Squared $\ell_2$ objective 
& \checkmark 
& \checkmark \\

Robust $\ell_1$ objective 
& Not considered 
& \checkmark \\

Solution guarantee 
& Global optimum of \eqref{LSformulation} 
& $\epsilon$-critical point of \eqref{Penalty} \\

Computational complexity 
& $O(n^2)$ 
& $O(n^2)$ \\

Scalability to larger networks 
& \checkmark 
& \checkmark \\
\hline
\end{tabular}
\end{table}

{
The effects of these differences are further evaluated in the following subsections through baseline comparisons, box-constraint analysis, and numerical tests under different target locations, beamwidths, and network sizes.
}

    \subsection{Baseline comparison and box analysis}
    \subsubsection{Comparison with $\mathrm{EDMR}_p$}
    \begin{figure*}[!t]
    \centering
    \subfloat[RMSE versus SNR$_0$]{%
        \includegraphics[width=0.32\textwidth]{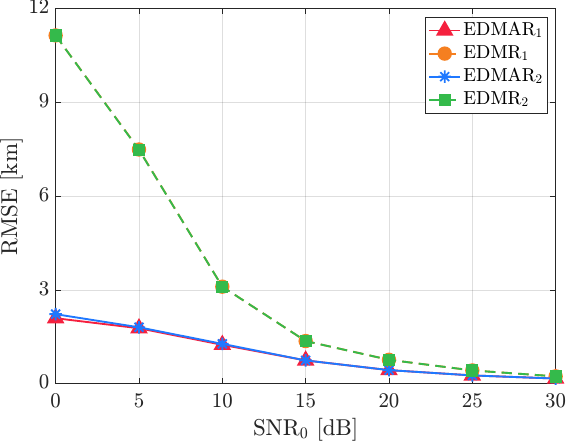}%
        \label{fig:MPA1rmse}%
    }%
    \hfill 
    \subfloat[Time versus SNR$_0$]{%
        \includegraphics[width=0.32\textwidth]{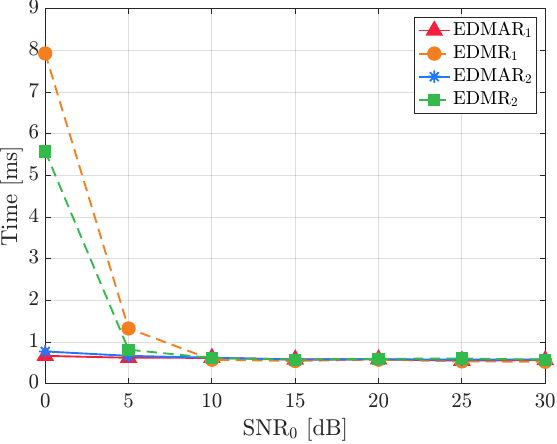}%
        \label{fig:MPA1time}%
    }%
    \hfill
    \subfloat[Eigenratio versus SNR$_0$]{%
        \includegraphics[width=0.32\textwidth]{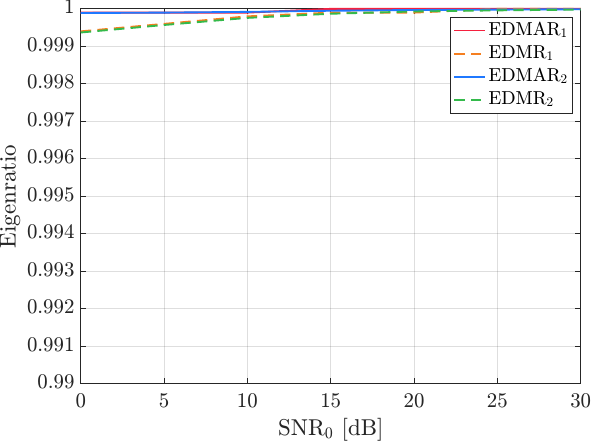}%
        \label{fig:MPA1Eigenratio}%
    }%
    \caption{Comparison between EDMAR$_p$ and EDMR$_p$, when $(\theta, \phi) = (6.9^\circ, 4.9^\circ)$ and $(\overline{\theta}, \overline{\phi}) = (7^\circ, 5^\circ)$.}
    \label{fig:MPA1}
\end{figure*}

	To show the importance of the angle information, we consider the specific scenario for the target position at $(\theta, \phi) = (6.9^\circ, 4.9^\circ)$ under fixed beamwidths $(\overline{\theta}, \overline{\phi}) = (7^\circ, 5^\circ)$. 
    The performance of EDMAR$_p$ and EDMR$_p$ is shown in Fig. \ref{fig:MPA1}.
    Fig. \ref{fig:MPA1rmse} demonstrates that EDMAR$_1$ performs best and EDMAR$_p$ consistently achieves significantly higher localization accuracy than EDMR$_p$, particularly under high-noise conditions. Specifically, at $\mathrm{SNR}_{0} = 0\mathrm{~dB}$, the RMSE of EDMAR$_1$ and EDMAR$_2$ is approximately 2.10$\mathrm{~km}$ and 2.23$\mathrm{~km}$, respectively, whereas that of EDMR$_p$ exceeds 10$\mathrm{~km}$. This result clearly indicates that the angle constraints in EDMAR$_p$ effectively leverage prior angle information, thereby substantially improving localization precision under severe noise interference.

    Furthermore, EDMAR$_p$ maintains remarkable computational efficiency with an average execution time of 0.6$\mathrm{~ms}$ per Monte Carlo trial. Fig.~\ref{fig:MPA1Eigenratio} shows that the EDM quality obtained through the MM framework remains exceptionally high, with all Eigenratio values consistently above 99.9\%, which confirms that the penalty function precisely enforces the constraints without over-relaxation, thereby validating the robustness of the proposed optimization framework in preserving the geometric structure of the localization problem while efficiently incorporating both range and angle measurements.

     {
    \subsubsection{Tightness analysis of the proposed box constraints}

To quantify the tightness of the box constraints computed by Algorithm~\ref{alg:ComputeLU}, we consider the noise-free setting with different beamwidths and  evaluate the average relative box width
$
W_{\rm rel}:=\frac{1}{m}\sum_{i=1}^m \frac{u_i-l_i}{\|\bfx-\bfx_i\|^2}.
$

The corresponding results are shown in Fig.~\ref{fig:box_tightness}. In all tested cases, the coverage rate of the true squared distances,
$
\frac{1}{m}\sum_{i=1}^m \mathbf{1}\!\left\{\, l_i \le \|\bfx-\bfx_i\|^2 \le u_i \,\right\},
$
is equal to one. It can be seen that the proposed box bounds remain tight in relative terms. Moreover, when the beamwidth becomes wider, the corresponding box becomes looser, as reflected by the increase of $W_{\rm rel}$ from about 0.023 to about 0.032. This confirms that the proposed box construction effectively captures the strength of the available angle prior information. For a fixed beamwidth, the box width also varies slightly with the target location. The performance of EDMAR$_p$ under different target locations and beamwidths will be shown in  Section~5.3.1 and 5.3.2, respectively.

\begin{figure}[t]
    \centering
    \includegraphics[width=\linewidth]{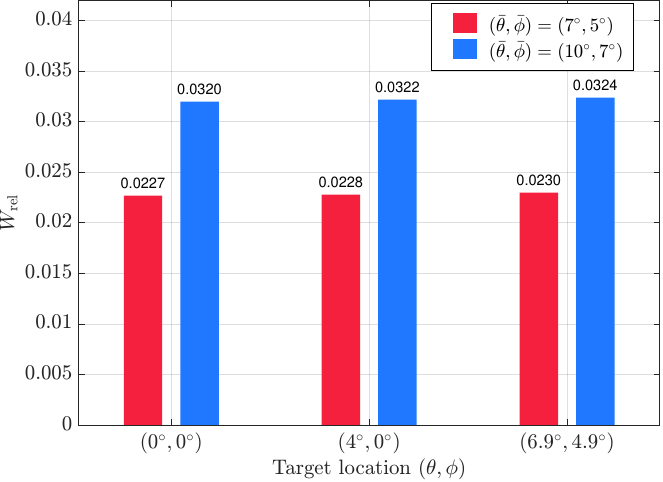}
    \caption{Average relative box width $W_{\rm rel}$ for three target locations under two beamwidth settings in the noise-free case.}
    \label{fig:box_tightness}
\end{figure}


}
    \subsection{Comparison with existing methods}
    \subsubsection{Performance under different target locations}
\begin{figure*}[!t]
    \centering
    \subfloat[$\theta = 0^{\circ}, \phi = 0^{\circ}$]{%
        \includegraphics[width=0.32\textwidth]{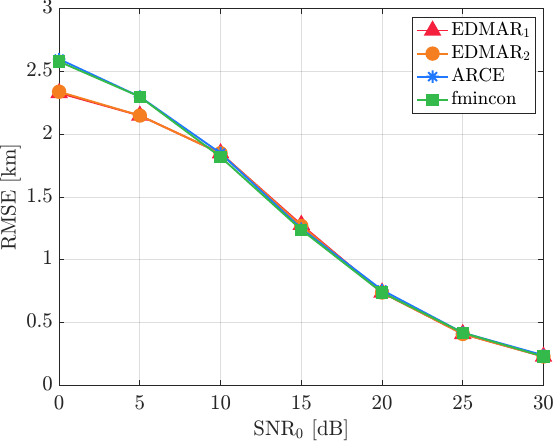}%
        \label{fig: xv1n5}%
    }%
    \hfill 
    \subfloat[$\theta = 4^{\circ}, \phi = 0^{\circ}$]{%
        \includegraphics[width=0.32\textwidth]{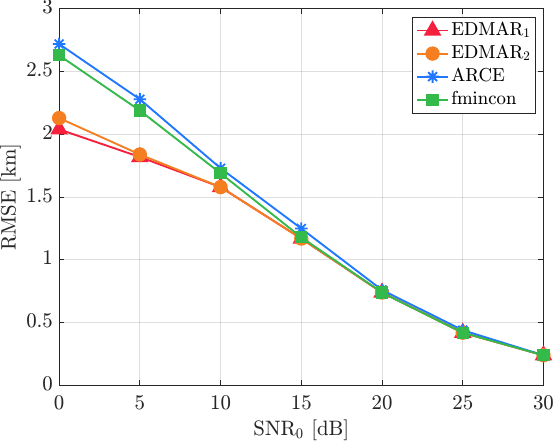}%
        \label{fig:xv2n5}%
    }%
    \hfill
    \subfloat[$\theta = 6.9^{\circ}, \phi = 4.9^{\circ}$]{%
        \includegraphics[width=0.32\textwidth]{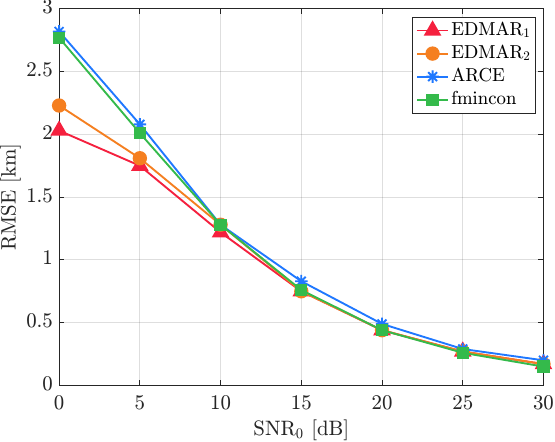}%
        \label{fig: xv3n5}%
    }%
    \caption{RMSE versus $\text{SNR}_0$ under different target locations, when $(\overline{\theta}, \overline{\phi}) = (7^\circ, 5^\circ)$.}
    \label{fig_m=5}
\end{figure*}
	    Different scenarios are analyzed, i.e., $\left(\theta, \phi\right)$ $\in$ $\left\{\left(0^{\circ}, 0^{\circ}\right),\left(4^{\circ}, 0^{\circ}\right),\left(6.9^{\circ}, 4.9^{\circ}\right)\right\}$.   
           Fig. \ref{fig_m=5} illustrates the RMSE versus $\mathrm{SNR}_{0}$ over a range of 0 $\mathrm{~dB}$ to 30 $\mathrm{~dB}$, { where each subfigure corresponds to a different target position} under fixed beamwidths  ($\overline{\theta} = 7^\circ$, $\overline{\phi} = 5^\circ$).
           The proposed EDMAR framework consistently achieves the lowest RMSE values across all configurations, while ARCE and fmincon exhibit competitive yet suboptimal performance.
           {
This demonstrates the effectiveness of the proposed EDM optimization framework that jointly exploits range and angle information for 3D SSLAR.
}
           Notably, \EDMAR$_1$ exhibits marginally better precision than \EDMAR$_2$ and ARCE, particularly under low SNR scenarios.
          {
This can be attributed to the robust $\ell_1$ objective.
At low SNR, the range measurements $\delta_{in}$ are more likely to exhibit large deviations, which may induce large residuals in the localization objective.
Unlike \EDMAR$_2$ and ARCE, which use squared $\ell_2$ least-squares criteria, \EDMAR$_1$ penalizes the distance residual $\sqrt{D_{in}}-\delta_{in}$ linearly, thereby reducing the influence of large residuals and improving robustness.
}
           {Furthermore, the localization performance of both \EDMAR$_p$\ algorithms varies with the target position. In particular, the results for $(\theta,\phi)=(6.9^\circ,4.9^\circ)$ are better than the others. This indicates that the effectiveness of the proposed method depends on the underlying target geometry. Such a result is also consistent with the box analysis in Section~5.2.2.}
           
    \begin{figure}[!t]
    \centering
    \includegraphics[width=\linewidth]{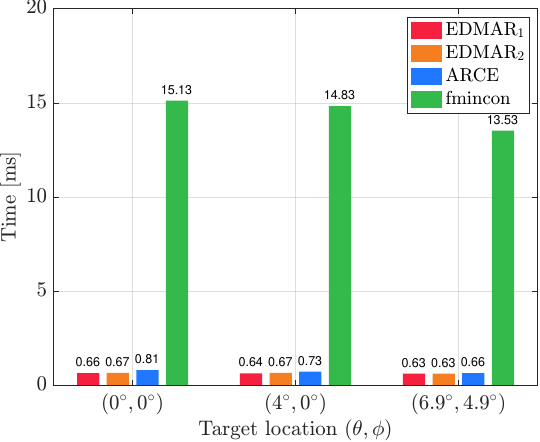}
    \caption{Average CPU time (ms) at $\mathrm{SNR}_0=10$ $\mathrm{~dB}$, when $(\overline{\theta}, \overline{\phi})=(7^\circ,5^\circ)$.}
    \label{fig:time_n5}
\end{figure}

{The computational efficiency of each algorithm is further compared in Fig.~\ref{fig:time_n5}, which reports the average CPU time (in milliseconds) at $\mathrm{SNR}_{0}=10$ dB for the three target locations.}
       EDMAR$_p$ runs relatively low and stable computation times across all scenarios.
	This is primarily because our implementation does not include the refinement step, which utilizes a heuristic gradient method to improve accuracy \citep{Re_2dBiswas_2005}. Instead, EDMAR$_p$ more effectively extracts angle and range information in the early stages, resulting in significantly enhanced localization accuracy without the additional time-consuming refinement step adopted by traditional EDM-based methods. 
	 The runtime of ARCE\footnote[3]{The implementations of ARCE and ROCE were reproduced based on the original descriptions. The runtime observed in our experiments appears faster than that reported in the original papers. This discrepancy may be due to differences in the hardware environment or variations in the implementation details.} is competitive while the MATLAB solver fmincon runs the slowest among all methods and is very time-consuming. 
    
    \subsubsection{Performance under different beamwidth settings}
    \begin{figure*}[!t]
    \centering
    \subfloat[$(\overline{\theta},\overline{\phi})=(7^\circ,5^\circ)$]{%
        \includegraphics[width=0.48\textwidth]{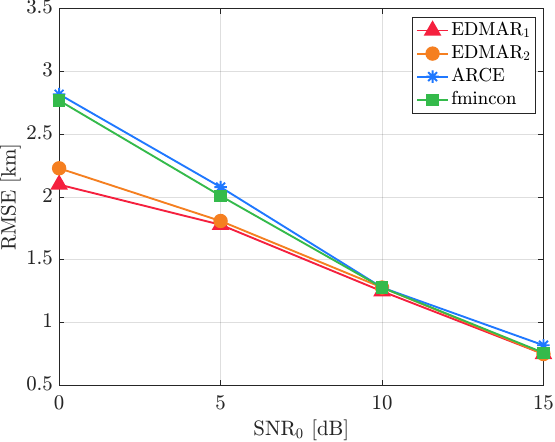}%
        \label{fig:rmse_bw75}%
    }%
    \hfill
    \subfloat[$(\overline{\theta},\overline{\phi})=(10^\circ,7^\circ)$]{%
        \includegraphics[width=0.48\textwidth]{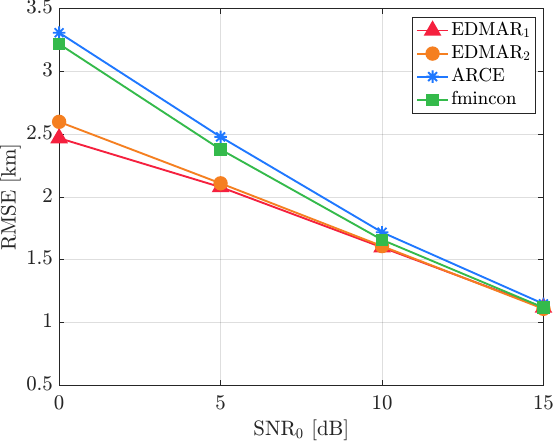}%
        \label{fig:rmse_bw107}%
    }%
    \caption{RMSE versus $\mathrm{SNR}_0$ for $(\theta,\phi)=(6.9^\circ,4.9^\circ)$ under different beamwidth settings.}
    \label{fig:rmse_beamwidth}
\end{figure*}
    Figure~\ref{fig:rmse_beamwidth} analyzes the RMSE of angle-constrained methods when $\theta = 6.9^\circ$, $\phi = 4.9^\circ$ with different main-lobe widths. Expanding beamwidths from $(\overline{\theta}=7^\circ, \overline{\phi}=5^\circ)$ to $(\overline{\theta}=10^\circ, \overline{\phi}=7^\circ)$ degrades the accuracy of all methods. Therefore, it is essential to give precise angle constraints to improve the localization accuracy of SSLAR. In practical scenarios, if the radar system can provide a more accurate bearing of the target, the localization result would be more precise. A proper strategy is to start with a wider beamwidth to guarantee that the target is detected, and then narrow the beamwidth to enhance localization accuracy. Despite this performance erosion, EDMAR$_p$ maintains superior robustness, still achieving a lower RMSE compared to the other methods. For example, EDMAR$_1$ achieves an RMSE of 2.47$\mathrm{~km}$ while   ARCE  3.31$\mathrm{~km}$ under the same noise condition $\mathrm{SNR} = 0\mathrm{~dB}$.
{
This improvement is consistent with the reduced sensitivity of the $\ell_1$ objective to large residuals compared with the squared $\ell_2$ least-squares formulation used in ARCE.
}
    Furthermore, EDMAR$_1$ leads to much smaller RMSE than EDMAR$_2$ as the $\mathrm{SNR}$ decreases, which means that EDMAR$_1$ is more robust than EDMAR$_2$.
    
     {
     \subsubsection{Performance under different network sizes}

We set
$
m\in\{5,10,20,50,75,100\},
$
with $(\theta,\phi)=(6.9^\circ,4.9^\circ)$, $(\overline{\theta},\overline{\phi})=(7^\circ,5^\circ)$, and $\mathrm{SNR}_0=0$ dB fixed. Since this setting goes beyond the benchmark geometry in \citep{Re_Enhance_a}, the receiver locations are generated randomly within the same spatial scale as the original test problem, with the transmitter fixed at the origin. All methods are tested on the same generated geometry for each network size.

Fig.~\ref{fig:size_effect} shows that the RMSE of all methods generally decreases as $m$ increases. Across all tested sizes, \EDMAR$_1$ and \EDMAR$_2$ consistently outperform ARCE and fmincon, with \EDMAR$_1$ giving the best overall accuracy.
However, the runtime of \EDMAR$_p$ grows much faster than that of ARCE and fmincon. This is mainly due to the truncated eigendecomposition required in the spectral projection step for the matrix-based MM updates.
Overall, these results show that the proposed EDM framework remains
numerically stable and effective as the network size increases,
 while improving its computational efficiency for larger-scale problems remains an important topic for future work.
     \begin{figure*}[!t]
    \centering
    \includegraphics[width=0.48\textwidth]{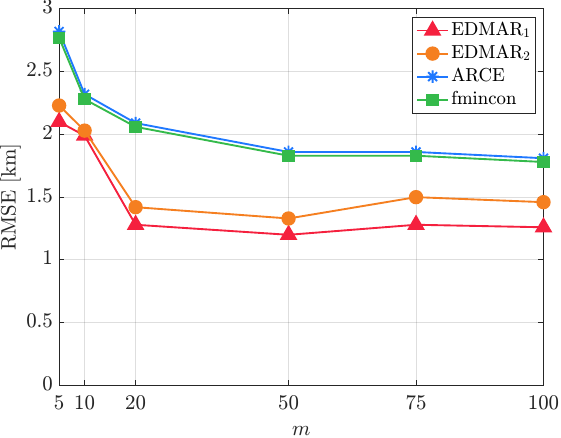}
    \hfill
    \includegraphics[width=0.48\textwidth]{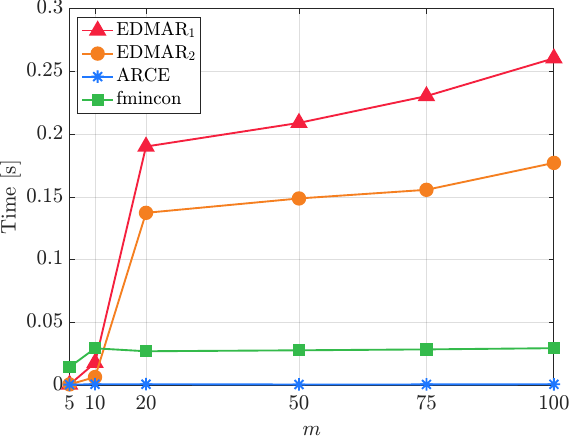}
    \caption{Performance under different network sizes, when $(\theta,\phi)=(6.9^\circ,4.9^\circ)$, $(\overline{\theta},\overline{\phi})=(7^\circ,5^\circ)$ and $\mathrm{SNR}_0=0$ dB.
    }
    \label{fig:size_effect}
\end{figure*}
}
\section{Conclusion} \label{Section-Conclusion}

    In this paper, we introduced a robust EDM optimization approach to address the 3D SSLAR in MPRNs. The proposed model reformulated the range and angle constraints as box constraints by deriving upper and lower bounds through solving a set of 2D constrained optimization subproblems, effectively handling the nonconvexity and geometrical information in SSLAR.
    
    To solve the resulting model, we designed a deterministic multi-start initialization strategy and applied the majorization penalty approach. Extensive numerical experiments validate the superiority of the proposed algorithms, denoted by EDMAR$_p$ ($p = 1, 2$). By leveraging the angle information, EDMAR$_p$ achieves notable improvements in localization accuracy compared with the baseline EDMR$_p$, while maintaining high computational efficiency. In comparison with state-of-the-art algorithms such as ARCE and the MATLAB solver fmincon, EDMAR$_p$ consistently yields lower RMSE, particularly in low SNR conditions. In addition to enhanced accuracy, EDMAR$_p$ demonstrates fast convergence, making it suitable for real-time applications in dynamic environments. 
    
    Finally, by incorporating box constraints derived from range and angle information, the proposed EDM-based model exhibits strong adaptability to complex radar configurations, ensuring broad applicability across diverse scenarios.

    {Future work will focus on extending the proposed framework to multi-target localization and tracking. In principle, such an extension can be formulated through a larger Euclidean distance matrix with partially unknown entries, which naturally leads to a matrix completion problem. From this viewpoint, the EDM methodology remains applicable. A major challenge is how to further exploit the underlying low-rank structure of the resulting matrix so as to improve both computational efficiency and localization accuracy. Future work will therefore focus on low-rank modeling, matrix completion, and scalable algorithm design within the proposed EDM framework using angle and range measurements.}
  

    \section*{CRediT authorship contribution statement}

    \textbf{Mingyu Zhao:} Conceptualization, Methodology, Software, Validation, Formal analysis, Data curation, Writing -- original draft, Visualization. 
    \textbf{Qingna Li:} Methodology, Validation, Resources, Writing -- review \& editing, Supervision, Funding acquisition. 
    \textbf{Hou-Duo Qi:} Conceptualization, Methodology, Validation, Resources, Writing -- review \& editing, Supervision, Funding acquisition.

    \section*{Funding}
    This work was supported by the National Natural Science Foundation of China [grant numbers: 12071036, 12271526], 
	and by the Hong Kong RGC General Research Fund [grant number: PolyU/15303124], and PolyU AMA projects [grant numbers: P0044200, P0045347].

    \section*{Declaration of competing interest}
    The authors declare that they have no known competing financial interests or personal relationships that could have appeared to influence the work reported in this paper.

\appendix


 \section{Proof of Lemma \ref{lem:extremum}} \label{Appendix-Lemma} 
 
 \begin{proof}
		First, by the definitions of $\theta$ and $\phi$ in \eqref{eq_twoangle} as well as $x >0$, it holds that
		\begin{equation}\label{xyz}
				y = x \tan \theta, \
				z = x \tan \phi, \
				x > 0,
		\end{equation}
		where $\theta$ and $\phi$ satisfy condition \eqref{angle1}.
		
		Since the equation $x^{2} + y^{2} + z^{2} = \|\mathbf{x}\|^{2}$ holds, together with \eqref{xyz}, we can represent  $x$, $y$ and $z$ in terms of $v_{1}$ and $v_{2}$ as follows
\[ 
     	x = \dfrac{\|\mathbf{x}\|}{\sqrt{1 + v_{1}^{2} + v_{2}^{2}}}, \
    	y = \dfrac{\|\mathbf{x}\| \, v_{1}}{\sqrt{1 + v_{1}^{2} + v_{2}^{2}}}, \
    	z = \dfrac{\|\mathbf{x}\| \, v_{2}}{\sqrt{1 + v_{1}^{2} + v_{2}^{2}}}.
\] 
		Substituting the above into $\|\mathbf{x}_{i} - \mathbf{x}\|^{2}$ yields
		$$
		\begin{array}{ll}
	&\|\mathbf{x}_{i} - \mathbf{x}\|^{2} \\
        = & (x_{i} - x)^{2} + (y_{i} - y)^{2} + (z_{i} - z)^{2} \\
			= & -2 \dfrac{ \|\mathbf{x}\|}{\sqrt{1 + v_{1}^{2} + v_{2}^{2}}} \left(x_{i} + y_{i} v_{1} + z_{i} v_{2}\right) + \|\mathbf{x}_i\|^{2} +\|\mathbf{x}\|^{2}. 
		\end{array}
		$$
        
        Recall the range constraint $\|\mathbf{x}\|^{2}=b^{2}$ and $\|\mathbf{x}_{i}\|^{2}=\Delta_{1i}$. 
        {Since $\bfx_1 =0$},
		 we can rewrite the above formula as
		\[
		h_{i}\left( \boldsymbol{v}  \right) = -2 \dfrac{b}{\sqrt{1 + v_{1}^{2} + v_{2}^{2}}} \left(x_{i} + y_{i} v_{1} + z_{i} v_{2}\right) + \Delta_{1i} +b^{2}.
		\] 
		This gives Equation \eqref{eq:h_{i}}. By simple calculation, we derive  \eqref{eq:partial_h_{i}}. The proof is finished. 
	\end{proof}

\section{Proof of Theorem \ref{th_candidateset}} \label{Appendix-Thm}

\begin{proof}
			We start by rewriting problem \eqref{eq_minf} as
			\begin{equation}\label{eq_minf1}
				\begin{array}{rl}
					\underset{\mathbf{v} \in \mathbb{R}^2}{\min} & h_{i}\left(\mathbf{v}\right) \\
					\text{s.t.} & v_{1} + \gamma_{a} \geq 0,\\
					& - v_{1} + \gamma_{a}\geq 0, \\
					& v_{2} + \gamma_{e} \geq 0, \\
					& - v_{2} + \gamma_{e} \geq 0.
				\end{array}
			\end{equation}
			Let $\mathbf{\mu} := \left(\mu_1, \mu_2, \mu_3, \mu_4\right) ^{\top}$  be the corresponding Lagrange multipliers in \eqref{eq_minf1}. The Lagrangian function of \eqref{eq_minf1} is
			\[
            \begin{aligned}
                \mathcal{L}_i (\mathbf{v}, \mathbf{\mu}) = h_{i}\left(\mathbf{v}\right) - \mu_1 (v_{1} + \gamma_{a}) - \mu_2 (- v_{1} + \gamma_{a})\\ - \mu_3 (v_{2} + \gamma_{e}) - \mu_4 (- v_{2} + \gamma_{e}).
            \end{aligned}
			\]
			Let $(\mathbf{v}, \mathbf{\mu})^{\top}$ be a KKT point satisfying the following conditions:
			\begin{subequations}\label{eq:KKT}
				\begin{numcases}{}
						\frac{\partial \mathcal{L}_i}{\partial v_1} (\mathbf{v}, \mathbf{\mu}) = \frac{\partial h_{i}}{\partial v_1} (\mathbf{v}) - \mu_1 + \mu_2 = 0, \label{eq:21a} \\
					\frac{\partial \mathcal{L}_i}{\partial v_2} (\mathbf{v}, \mathbf{\mu})  = \frac{\partial h_{i}}{\partial v_2} (\mathbf{v}) - \mu_3 + \mu_4 = 0, \label{eq:21b} \\
                    \begin{aligned}
                          v_1 + \gamma_a \geq 0, \quad - v_1 +  \gamma_a \geq 0, \\
					v_2 + \gamma_e \geq 0, \quad  - v_2 + \gamma_e \geq 0, 
                    \end{aligned}
   \label{eq:21c} \\
                    \begin{aligned}
                        \mu_1 (v_1 + \gamma_a) = 0, \quad \mu_2 ( - v_1 +\gamma_a) = 0, \\
					\mu_3 (v_2 + \gamma_e) = 0, \quad \mu_4 (- v_2+ \gamma_e) = 0, 
                    \end{aligned}
					\label{eq:21d} \\
					\mu_1 \geq 0, \quad \mu_2 \geq 0, \quad \mu_3 \geq 0, \quad \mu_4 \geq 0. \label{eq:21e}
				\end{numcases}
			\end{subequations}
			 Next, we proceed with a case-by-case analysis.\\
		    \textbf{Case 1.}
				If $\mu_1 = \mu_2 = \mu_3 = \mu_4 = 0$, then from \eqref{eq:21a} and \eqref{eq:21b}, it holds that
				$$
					\nabla h_{i}(\mathbf{v}) = \mathbf{0}.
				$$
				By Lemma \ref{lem:extremum}, it is equivalent to
				\begin{equation*} 
					\left\{
					\begin{aligned}
						& \dfrac{v_{1}\left(x_{i} + y_{i} v_{1} + z_{i} v_{2}\right) - y_i \left(1 + v_1^2 + v_2^2\right)}
						{\left(1 + v_{1}^{2} + v_{2}^{2}\right)^{\frac{3}{2}}} = 0, \\
						& \dfrac{v_{2} \left(x_{i} + y_{i} v_{1} + z_{i} v_{2}\right) - z_i \left(1 + v_1^2 + v_2^2\right)}
						{\left(1 + v_{1}^{2} + v_{2}^{2}\right)^{\frac{3}{2}}}  = 0.
					\end{aligned}
					\right.
				\end{equation*}
				One can obtain that
				$\mathbf{v} = \left(\frac{y_{i}}{x_{i}},  \frac{z_{i}}{x_{i}}\right) ^{\top}.
					$
					\\
				Therefore, if $ -\gamma_{a} \leq \frac{y_{i}}{x_{i}} \leq \gamma_{a} \text{ and }  -\gamma_{e} \leq \frac{z_{i}}{x_{i}} \leq \gamma_{e}$, then $\left(\frac{y_{i}}{x_{i}}, \frac{z_{i}}{x_{i}}, 0, 0, 0, 0\right)^{\top}$ satisfies the KKT conditions \eqref{eq:KKT}.\\
			 \textbf{Case 2. Boundary Solutions for One Active Constraint.}\\
				\textbf{Case 2.1.}  $\mu_1 > 0$, $\mu_2 = \mu_3 = \mu_4 = 0$.\\ 
				From the complementary slackness condition \eqref{eq:21d}, we have
				$
				v_{1} = -\gamma_{a}.
				$
				Substituting $\mu_2 = \mu_3 = \mu_4 = 0$ into the KKT condition \eqref{eq:21b}, we obtain 
                \begin{equation*}
                    \begin{aligned}
                        \frac{\partial h_{i}}{\partial v_2}\left(-\gamma_{a}, v_2\right) = \frac{2 b v_2}{\left(1 + \gamma_{a}^{2} + v_2^{2}\right)^{\frac{3}{2}}} \left(x_{i} - \gamma_{a} y_{i} + z_{i} v_2\right) \\
                        - \frac{2 b z_{i}}{\sqrt{1 + \gamma_{a}^{2} + v_2^{2}}} = 0, 
                    \end{aligned}
                \end{equation*}
                which gives $ v_2 = \frac{\left(1 + \gamma_{a}^{2}\right) z_{i}}{x_{i} - \gamma_{a} y_{i}}.
				$
				By \eqref{eq:21a}, one can obtain that 
    \begin{equation*}
        \begin{aligned}
            \mu_1 =& \dfrac{\partial h_{i}}{\partial v_{1}}\left(-\gamma_{a}, \frac{\left(1 + \gamma_{a}^{2}\right) z_{i}}{x_{i} - \gamma_{a} y_{i}}\right) \\ =&
				\frac{-2 b\left(\left(x_{i}-\gamma_{a} y_{i}\right)^{2}+\left(1+\gamma_{a}^{2}\right) z_{i}^{2}\right)}{Q_1^{\frac{3}{2}}\left(x_{i}-\gamma_{a} y_{i}\right)^{2}}\left(\gamma_{a} x_{i}+y_{i}\right),
        \end{aligned}
    \end{equation*}
where $Q_1 := 1+\gamma_{a}^{2}+\frac{\left(1+\gamma_{a}^{2}\right)^{2} z_{i}^{2}}{\left(x_{i}-\gamma_{a} y_{i}\right)^{2}}$.\\
	Notice that $\mu_1 > 0$ is equivalent to $\gamma_{a}x_i + y_i <  0$.
				Therefore, if $-\gamma_{e} \leq \frac{\left(1 + \gamma_{a}^{2}\right) z_{i}}{x_{i} - \gamma_{a} y_{i}} \leq \gamma_{e}$ and $\gamma_{a}x_i + y_i <  0$, then 
				$ \left(-\gamma_{a}, \frac{\left(1 + \gamma_{a}^{2}\right) z_{i}}{x_{i} - \gamma_{a} y_{i}}, \mu_1, 0, 0, 0\right) ^{\top}$ is a solution of the KKT system \eqref{eq:KKT}.\\
			 Similarly,  we can discuss in the same way and obtain the following results.\\
					 \textbf{Case 2.2.} 	$\mu_2 > 0$, $\mu_1 = \mu_3 = \mu_4 = 0.$ \\
                     If $-\gamma_{e} \leq \frac{\left(1 + \gamma_{a}^{2}\right) z_{i}}{x_{i} + \gamma_{a} y_{i}} \leq \gamma_{e}$ and $\gamma_{a}x_i - y_i < 0$, then 
						$\left(\gamma_{a},  \frac{\left(1 + \gamma_{a}^{2}\right) z_{i}}{x_{i} + \gamma_{a} y_{i}}, 0, \mu_2, 0, 0\right) ^{\top}$ is a solution of the KKT system \eqref{eq:KKT}, where $\mu_2 = \frac{-2 b\left(\left(x_{i}+\gamma_{a} y_{i}\right)^{2}+\left(1+\gamma_{a}^{2}\right) z_{i}^{2}\right)}{Q_2^{\frac{3}{2}}\left(x_{i}+\gamma_{a} y_{i}\right)^{2}}\left(\gamma_{a} x_{i}-y_{i}\right)$ 
						and $Q_2 := 1+\gamma_{a}^{2}+\frac{\left(1+\gamma_{a}^{2}\right)^{2} z_{i}^{2}}{\left(x_{i}+\gamma_{a} y_{i}\right)^{2}}$.
					\\ \textbf{Case 2.3.} $\mu_3 > 0$, $\mu_1 = \mu_2 = \mu_4 = 0.$\\  If $-\gamma_{a} \leq \frac{\left(1 +\gamma_{e}^{2}\right) y_{i}}{x_{i} - \gamma_{e} z_{i}} \leq \gamma_{a}$ and $\gamma_{e} x_{i}+z_{i} < 0$, then $\left(\frac{\left(1+\gamma_{e}^{2}\right) y_{i}}{x_{i} - \gamma_{e} z_{i}}, -\gamma_{e}, 0, 0, \mu_3 , 0\right)^{\top}$  is a solution of the KKT system \eqref{eq:KKT}, where $\mu_3 = \frac{-2 b\left(\left(x_{i} - \gamma_{e}z_{i}\right)^{2}+\left(1+\gamma_{e}^{2}\right) y_{i}^{2}\right)}{Q_3^{\frac{3}{2}}\left(x_{i} - \gamma_{e} z_{i}\right)^{2}}\left(\gamma_{e} x_{i}+z_{i}\right)
					$ and $Q_3 := 1+\gamma_{e}^{2}+\frac{\left(1+\gamma_{e}^{2}\right)^{2} y_{i}^{2}}{\left(x_{i} - \gamma_{e} z_{i}\right)^{2}}$.
					\\ \textbf{Case 2.4.}  $\mu_4 > 0$, $\mu_1 = \mu_2 = \mu_3 = 0.$\\ If $-\gamma_{a} \leq \frac{\left(1 + \gamma_{e}^{2}\right) y_{i}}{x_{i} + \gamma_{e} z_{i}} \leq \gamma_{a}$ and $\gamma_{e} x_{i} - z_{i} < 0$, then 
					$\left(\frac{\left(1+\gamma_{e}^{2}\right) y_{i}}{x_{i} + \gamma_{e} z_{i}}, \gamma_{e}, 0, 0, 0, \mu_4\right)^{\top}$ is a solution of the KKT system \eqref{eq:KKT}, where $\mu_4 = \frac{-2 b\left(\left(x_{i} + \gamma_{e} z_{i}\right)^{2}+\left(1+\gamma_{e}^{2}\right) y_{i}^{2}\right)}{Q_4^{\frac{3}{2}}\left(x_{i} + \gamma_{e} z_{i}\right)^{2}}\left(\gamma_{e} x_{i} - z_{i}\right)$ 
					and $Q_4 := 1+\gamma_{e}^{2}+\frac{\left(1+\gamma_{e}^{2}\right)^{2} y_{i}^{2}}{\left(x_{i} + \gamma_{e} z_{i}\right)^{2}}$.
			\\			
			 \textbf{Case 3. Corner Solutions for Two Active Constraints.}\\
				\textbf{Case 3.1.}  $\mu_1 > 0$, $\mu_3 > 0$, $\mu_2 = \mu_4 = 0$. \\
				From the complementary slackness conditions \eqref{eq:21d}, we have
				$$
				v_{1} + \gamma_{a} = 0 \quad \text{and} \quad v_{2} + \gamma_{e} = 0,
				$$
				implying that $\mathbf{v} = \left(-\gamma_{a}, -\gamma_{e}\right) ^{\top}$.
				By \eqref{eq:21a} and \eqref{eq:21b}, we get 
				$$\mu_1 = \dfrac{\partial h_{i}}{\partial v_{1}}\left(-\gamma_{a}, -\gamma_{e}\right) = -\dfrac{2 b \left(\gamma_{a} x_{i} + y_{i} (1 + \gamma_{e}^2) - \gamma_{a} \gamma_{e} z_{i}\right)}{\left(1 + \gamma_{a}^{2} + \gamma_{e}^{2}\right)^{\frac{3}{2}}},$$ 
				$$\mu_3 = \dfrac{\partial h_{i}}{\partial v_{2}}\left(-\gamma_{a}, -\gamma_{e}\right) = -\dfrac{2 b \left(\gamma_{e} x_{i} + z_{i} (1 + \gamma_{a}^2) - \gamma_{a} \gamma_{e} y_{i}\right)}{\left(1 + \gamma_{a}^{2} + \gamma_{e}^{2}\right)^{\frac{3}{2}}}.$$ 
				Notice that $\mu_1 > 0$ is equivalent to $\gamma_{a} x_{i} + y_{i} (1 + \gamma_{e}^2) - \gamma_{a} \gamma_{e} z_{i} < 0$, and $\mu_3 > 0$ is equivalent to $\gamma_{e} x_{i} + z_{i} (1 + \gamma_{a}^2) - \gamma_{a} \gamma_{e} y_{i} < 0$.
				Therefore, if $\gamma_{a} x_{i} + y_{i} (1 + \gamma_{e}^2) - \gamma_{a} \gamma_{e} z_{i} < 0$ and  $\gamma_{e} x_{i} + z_{i} (1 + \gamma_{a}^2) - \gamma_{a} \gamma_{e} y_{i} <  0$, then
				$\left(-\gamma_{a}, -\gamma_{e},  \mu_1, 0, \mu_3 , 0\right) ^{\top}$
				is a solution of the KKT system \eqref{eq:KKT}.\\
			    Similarly, we can discuss in the same way and obtain the following results.
					\\  \textbf{Case 3.2.} $\mu_1 > 0$, $\mu_4 > 0$, $\mu_2 = \mu_3 = 0.$\\ If $\gamma_{a} x_{i} + y_{i} (1 + \gamma_{e}^2) - \gamma_{a} \gamma_{e} z_{i} < 0$ and $\gamma_{e} x_{i} + z_{i} (1 + \gamma_{a}^2) - \gamma_{a} \gamma_{e} y_{i} < 0$, then $\left(-\gamma_{a}, \gamma_{e}, \mu_1 , 0, 0, \mu_4\right) ^{\top}$  is a solution of the KKT system \eqref{eq:KKT}, where 
					$\mu_1 = -\frac{2 b \left(\gamma_{a} x_{i} + y_{i} (1 + \gamma_{e}^2) + \gamma_{a} \gamma_{e} z_{i}\right)}{\left(1 + \gamma_{a}^{2} + \gamma_{e}^{2}\right)^{\frac{3}{2}}}$ and
					$\mu_4 = -\frac{2 b \left(\gamma_{e} x_{i} - z_{i} (1 + \gamma_{a}^2) - \gamma_{a} \gamma_{e} y_{i}\right)}{\left(1 + \gamma_{a}^{2} + \gamma_{e}^{2}\right)^{\frac{3}{2}}}$.
					\\ \textbf{Case 3.3.}  $\mu_2 > 0$, $\mu_3 > 0$, $\mu_1 = \mu_4 = 0.$\\
					If $\gamma_{a} x_{i} - y_{i} (1 + \gamma_{e}^2) - \gamma_{a} \gamma_{e} z_{i} < 0$ and $\gamma_{e} x_{i} + z_{i} (1 + \gamma_{a}^2) + \gamma_{a} \gamma_{e} y_{i} < 0$, then $\left(\gamma_{a}, -\gamma_{e}, 0, \mu_2, \mu_3, 0\right) ^{\top}$ is a solution of the KKT system \eqref{eq:KKT}, where 
					$\mu_2 = -\frac{2 b \left(\gamma_{a} x_{i} - y_{i} (1 + \gamma_{e}^2) - \gamma_{a} \gamma_{e} z_{i}\right)}{\left(1 + \gamma_{a}^{2} + \gamma_{e}^{2}\right)^{\frac{3}{2}}}$ and $\mu_3 = -\frac{2 b \left(\gamma_{e} x_{i} + z_{i} (1 + \gamma_{a}^2) + \gamma_{a} \gamma_{e} y_{i}\right)}{\left(1 + \gamma_{a}^{2} + \gamma_{e}^{2}\right)^{\frac{3}{2}}}$.
					\\ \textbf{Case 3.4.}  $\mu_2 > 0$, $\mu_4 > 0$, $\mu_1 = \mu_3 = 0.$\\
					If $\gamma_{a} x_{i} - y_{i} (1 + \gamma_{e}^2) + \gamma_{a} \gamma_{e} z_{i} < 0$ and $\gamma_{e} x_{i} - z_{i} (1 + \gamma_{a}^2) + \gamma_{a} \gamma_{e} y_{i} < 0$, then $\left(\gamma_{a}, \gamma_{e}, 0, \mu_2, 0, \mu_4\right) ^{\top}$ is a solution of the KKT system \eqref{eq:KKT}, where
					$\mu_2  =  -\frac{2 b \left(\gamma_{a} x_{i} - y_{i} (1 + \gamma_{e}^2) + \gamma_{a} \gamma_{e} z_{i}\right)}{\left(1 + \gamma_{a}^{2} + \gamma_{e}^{2}\right)^{\frac{3}{2}}}$ and $\mu_4 = -\frac{2 b \left(\gamma_{e} x_{i} - z_{i} (1 + \gamma_{a}^2) + \gamma_{a} \gamma_{e} y_{i}\right)}{\left(1 + \gamma_{a}^{2} + \gamma_{e}^{2}\right)^{\frac{3}{2}}}$.
			
			Note that the linear independent constraint qualification (LICQ) holds automatically for \eqref{eq_minf}. Therefore, for any local minimizer $\mathbf{v}$ of \eqref{eq_minf}, there is unique Lagrangian multiplier $\mathbf{\mu}$ such that $(\mathbf{v}, \mathbf{\mu})^{\top}$ is the solution of KKT system \eqref{eq:KKT}. In the same way, one can obtain the candidate KKT solutions for  \eqref{eq_maxf}. The complete set of candidate KKT solutions for both  \eqref{eq_minf} and \eqref{eq_maxf} is 
			summarized in Table \ref{ta_candidate_L}. Moreover, since the feasible set $V$ is a closed and bounded box in $\mathbb{R}^2$, and $h_i(\mathbf{v})$ is continuously differentiable on $V$, the Weierstrass theorem ensures that both the global minimum and maximum are attained. By the optimality theory, any local extremum must satisfy the KKT conditions. Therefore, evaluating $h_i(\mathbf{v})$ over the finite candidate set in Table \ref{ta_candidate_L} and selecting the best value yields the global solution of problems \eqref{eq_minf} and \eqref{eq_maxf}.
		\end{proof}


 \bibliographystyle{v3elsarticle-num-names} 
 \bibliography{v3sslarRef}

@article{Re_MPRNs_2,
  author       = {Brambilla, Mattia and Gaglione, Domenico and Soldi, Giovanni and Mendrzik, Rico and Ferri, Gabriele and LePage, Kevin D. and Nicoli, Monica and Willett, Peter and Braca, Paolo and Win, Moe Z.},
  title        = {Cooperative Localization and Multitarget Tracking in Agent Networks With the Sum-Product Algorithm},
  journal      = {IEEE Open J. Signal Process.},
  booktitle    = {},
  volume       = {3},
  number       = {},
  pages        = {169--195},
  year         = {2022},
  month        = {},
  publisher    = {IEEE},
  organization = {},
  doi          = {10.1109/OJSP.2022.3154684}
}

@article{Re_MPRNs_3,
  author       = {Aubry, Augusto and Braca, Paolo and De Maio, Antonio and Marino, Angela},
  title        = {2-{D} {PBR} Localization Complying With Constraints Forced by Active Radar Measurements},
  journal      = {IEEE Trans. Aerosp. Electron. Syst.},
  booktitle    = {},
  volume       = {57},
  number       = {5},
  pages        = {2647--2660},
  year         = {2021},
  month        = {},
  publisher    = {IEEE},
  organization = {},
  doi          = {}
}

@incollection{Re_multibetter,
  author       = {O’Hagan, Daniel W. and Doughty, Shaun R. and Inggs, Michael R.},
  title        = {Multistatic Radar Systems},
  journal      = {},
  booktitle    = {Academic Press Library in Signal Processing},
  volume       = {7},
  number       = {},
  pages        = {253--275},
  year         = {2018},
  month        = {},
  publisher    = {Elsevier},
  organization = {},
  doi          = {}
}

@article{Re_Enhance_a,
  author       = {Aubry, Augusto and Braca, Paolo and De Maio, Antonio and Marino, Angela},
  title        = {Enhanced Target Localization With Deployable Multiplatform Radar Nodes Based on Non-Convex Constrained Least Squares Optimization},
  journal      = {IEEE Trans. Signal Process.},
  booktitle    = {},
  volume       = {70},
  number       = {},
  pages        = {1282--1294},
  year         = {2022},
  month        = {},
  publisher    = {IEEE},
  organization = {},
  doi          = {10.1109/TSP.2022.3147037}
}

@article{Re_multibetter_1,
  author       = {Charlish, Alexander and Nadjiasngar, Roaldje and Klemm, R.},
  title        = {Sensor Management for Radar Networks},
  journal      = {Novel Radar Techniques and Applications: Waveform Diversity and Cognitive Radar and Target Tracking and Data Fusion},
  booktitle    = {},
  volume       = {2},
  number       = {},
  pages        = {457--488},
  year         = {2017},
  month        = {},
  publisher    = {IET London, UK},
  organization = {},
  doi          = {}
}

@article{Re_2dAubry_2020,
  author       = {Aubry, Augusto and Carotenuto, Vincenzo and De Maio, Antonio and Pallotta, Luca},
  title        = {Localization in {2D} {PBR} With Multiple Transmitters of Opportunity: A Constrained Least Squares Approach},
  journal      = {IEEE Trans. Signal Process.},
  booktitle    = {},
  volume       = {68},
  number       = {},
  pages        = {634--646},
  year         = {2020},
  month        = {},
  publisher    = {IEEE},
  organization = {},
  doi          = {10.1109/TSP.2020.2964235}
}

@inproceedings{Re_2dBiswas_2005,
  author       = {Biswas, Pratik and Aghajan, Hamid and Ye, Yin Yu},
  title        = {Semidefinite Programming Algorithms for Sensor Network Localization Using Angle Information},
  journal      = {},
  booktitle    = {Conf. Rec. 39th Asilomar Conf. Signals, Systems and Computers},
  volume       = {},
  number       = {},
  pages        = {220--224},
  year         = {2005},
  month        = {},
  publisher    = {},
  organization = {IEEE},
  doi          = {10.1109/ACSSC.2005.1599736}
}

@article{Re_Marino_2024,
  author       = {Marino, Angela and Soldi, Giovanni and Gaglione, Domenico and Aubry, Augusto and Braca, Paolo and De Maio, Antonio and Willett, Peter},
  title        = {3{D} Localization and Tracking Methods for Multiplatform Radar Networks},
  journal      = {IEEE Aerosp. Electron. Syst. Mag.},
  booktitle    = {},
  volume       = {39},
  number       = {5},
  pages        = {18--37},
  year         = {2024},
  month        = {},
  publisher    = {IEEE},
  organization = {},
  doi          = {10.1109/MAES.2024.3366139}
}

@article{Re_Jia_2025,
  author       = {Jia, Tian Yi and Ke, Xiao Chuan and Liu, Hong Wei and Ho, K. C. and Su, Hong Tao},
  title        = {Target Localization and Sensor Self-Calibration of Position and Synchronization by Range and Angle Measurements},
  journal      = {IEEE Trans. Signal Process.},
  booktitle    = {},
  volume       = {73},
  number       = {},
  pages        = {340--355},
  year         = {2025},
  month        = {},
  publisher    = {IEEE},
  organization = {},
  doi          = {10.1109/TSP.2024.3520909}
}

@article{Re_CWLS_2006,
  author       = {Cheung, Ka Wai and So, Hing-Cheung and Ma, Wing-Kin and Chan, Yiu-Tong},
  title        = {A constrained least squares approach to mobile positioning: algorithms and optimality},
  journal      = {EURASIP J. Adv. Signal Process.},
  booktitle    = {},
  volume       = {2006},
  number       = {1},
  pages        = {1--23},
  year         = {2006},
  month        = {},
  publisher    = {Springer},
  organization = {},
  doi          = {10.1155/ASP/2006/20858}
}

@article{Re_SWLS_2008,
  author       = {Beck, Amir and Teboulle, Marc and Chikishev, Zahar},
  title        = {Iterative Minimization Schemes for Solving the Single Source Localization Problem},
  journal      = {SIAM J. Optim.},
  booktitle    = {},
  volume       = {19},
  number       = {3},
  pages        = {1397--1416},
  year         = {2008},
  month        = {},
  publisher    = {SIAM},
  organization = {},
  doi          = {}
}

@article{Re_2SWLS_2016,
  author       = {Yang, Heeseong and Chun, Joohwan},
  title        = {An Improved Algebraic Solution for Moving Target Localization in Noncoherent {MIMO} Radar Systems},
  journal      = {IEEE Trans. Signal Process.},
  booktitle    = {},
  volume       = {64},
  number       = {1},
  pages        = {258--270},
  year         = {2016},
  month        = {},
  publisher    = {IEEE},
  organization = {},
  doi          = {10.1109/TSP.2015.2477803}
}

@inproceedings{Re_YeSDP_2004,
  author       = {Biswas, Pratik and Ye, Yin Yu},
  title        = {Semidefinite Programming for Ad Hoc Wireless Sensor Network Localization},
  journal      = {},
  booktitle    = {Proc. 3rd Int. Symp. Information Processing in Sensor Networks},
  volume       = {},
  number       = {},
  pages        = {46--54},
  year         = {2004},
  month        = {},
  publisher    = {},
  organization = {},
  doi          = {10.1145/984622.984630}
}

@article{Re_YeSDP_2006,
  author       = {Biswas, Pratik and Liang, T.-C. and Toh, K.-C. and Ye, Yin Yu and Wang, T.-C.},
  title        = {Semidefinite Programming Approaches for Sensor Network Localization With Noisy Distance Measurements},
  journal      = {IEEE Trans. Autom. Sci. Eng.},
  booktitle    = {},
  volume       = {3},
  number       = {4},
  pages        = {360--371},
  year         = {2006},
  month        = {},
  publisher    = {IEEE},
  organization = {},
  doi          = {10.1109/TASE.2006.877401}
}

@article{Re_QiSNT_2013,
  author       = {Qi, Hou-Duo},
  title        = {A Semismooth {Newton} Method for the Nearest {Euclidean} Distance Matrix Problem},
  journal      = {SIAM J. Matrix Anal. Appl.},
  booktitle    = {},
  volume       = {34},
  number       = {1},
  pages        = {67--93},
  year         = {2013},
  month        = {},
  publisher    = {SIAM},
  organization = {},
  doi          = {}
}

@article{Re_QiYuan_2014,
  author       = {Qi, Hou-Duo and Yuan, Xiao Ming},
  title        = {Computing the Nearest {Euclidean} Distance Matrix With Low Embedding Dimensions},
  journal      = {Math. Program.},
  booktitle    = {},
  volume       = {147},
  number       = {1},
  pages        = {351--389},
  year         = {2014},
  month        = {},
  publisher    = {Springer},
  organization = {},
  doi          = {}
}

@article{Re_QiLag_2013,
  author       = {Qi, Hou-Duo and Xiu, Nai Hua and Yuan, Xiao Ming},
  title        = {A {Lagrangian} Dual Approach to the Single-Source Localization Problem},
  journal      = {IEEE Trans. Signal Process.},
  booktitle    = {},
  volume       = {61},
  number       = {15},
  pages        = {3815--3826},
  year         = {2013},
  month        = {},
  publisher    = {IEEE},
  organization = {},
  doi          = {}
}

@article{Re_Zhoubox_2018,
  author       = {Zhou, Sheng Long and Xiu, Nai Hua and Qi, Hou-Duo},
  title        = {A Fast Matrix Majorization-Projection Method for Penalized Stress Minimization With Box Constraints},
  journal      = {IEEE Trans. Signal Process.},
  booktitle    = {},
  volume       = {66},
  number       = {16},
  pages        = {4331--4346},
  year         = {2018},
  month        = {},
  publisher    = {IEEE},
  organization = {},
  doi          = {}
}

@article{Re_Zhourobust_2018,
  author       = {Zhou, Sheng Long and Xiu, Nai Hua and Qi, Hou-Duo},
  title        = {Robust {Euclidean} Embedding via {EDM} Optimization},
  journal      = {Math. Program. Comput.},
  booktitle    = {},
  volume       = {12},
  number       = {3},
  pages        = {337--387},
  year         = {2020},
  month        = {},
  publisher    = {Springer},
  organization = {},
  doi          = {}
}

@article{Re_Shi_2023,
  author       = {Shi, He and Li, Qing Na},
  title        = {A Facial Reduction Approach for the Single Source Localization Problem},
  journal      = {J. Glob. Optim.},
  booktitle    = {},
  volume       = {87},
  number       = {2},
  pages        = {831--855},
  year         = {2023},
  month        = {},
  publisher    = {Springer},
  organization = {},
  doi          = {}
}

@article{Re_Geng,
  author       = {Geng, Zhe and Wang, Bei Ning and Yan, He and Zhang, Jin Dong and Zhu, Dai Yin},
  title        = {Moving Target Detection and Tracking With Multiplatform Radar Network ({MRN})},
  journal      = {IET Radar Sonar Navig.},
  booktitle    = {},
  volume       = {16},
  number       = {5},
  pages        = {815--824},
  year         = {2022},
  month        = {},
  publisher    = {Wiley},
  organization = {},
  doi          = {}
}

@article{Re_Brambilla,
  author       = {Brambilla, Mattia and Gaglione, Domenico and Soldi, Giovanni and Mendrzik, Rico and Ferri, Gabriele and LePage, Kevin D. and Nicoli, Monica and Willett, Peter and Braca, Paolo and Win, Moe Z.},
  title        = {Cooperative Localization and Multitarget Tracking in Agent Networks With the Sum-Product Algorithm},
  journal      = {IEEE Open J. Signal Process.},
  booktitle    = {},
  volume       = {3},
  number       = {},
  pages        = {169--195},
  year         = {2022},
  month        = {},
  publisher    = {IEEE},
  organization = {},
  doi          = {10.1109/OJSP.2022.3154684}
}

@article{Re_Sun_2017,
  author       = {Sun, Ying and Babu, Prabhu and Palomar, Daniel P.},
  title        = {Majorization-Minimization Algorithms in Signal Processing, Communications, and Machine Learning},
  journal      = {IEEE Trans. Signal Process.},
  booktitle    = {},
  volume       = {65},
  number       = {3},
  pages        = {794--816},
  year         = {2016},
  month        = {},
  publisher    = {IEEE},
  organization = {},
  doi          = {}
}

@book{Re_Numerical,
  author       = {Nocedal, Jorge and Wright, Stephen J.},
  title        = {Numerical Optimization},
  journal      = {},
  booktitle    = {},
  volume       = {},
  number       = {},
  pages        = {},
  year         = {2006},
  month        = {},
  publisher    = {Springer},
  organization = {},
  doi          = {}
}

@article{Re_Gower_1985,
  author       = {Gower, John Clifford},
  title        = {Properties of {Euclidean} and Non-{Euclidean} Distance Matrices},
  journal      = {Linear Algebra Appl.},
  booktitle    = {},
  volume       = {67},
  number       = {},
  pages        = {81--97},
  year         = {1985},
  month        = {},
  publisher    = {Elsevier},
  organization = {},
  doi          = {}
}

@article{Re_Wu_2023,
  author       = {Wu, Xiao Ping and Zhao, Li and Zhu, Xue Fen},
  title        = {Efficient Semidefinite Solutions for {TDOA}-Based Source Localization Under Unknown {PS}},
  journal      = {Pervasive Mobile Comput.},
  volume       = {91},
  number       = {},
  pages        = {},
  note         = {{A}rt. no. 101783},
  year         = {2023},
  month        = {Apr.},
  publisher    = {Elsevier},
  doi          = {10.1016/j.pmcj.2023.101783}
}

@incollection{Re_Marti_2018,
  author       = {Mart{\'i}, Rafael and Lozano, Jose A. and Mendiburu, Alexander and Hernando, Leticia},
  title        = {Multi-Start Methods},
  journal      = {},
  booktitle    = {Handbook of Heuristics},
  volume       = {},
  number       = {},
  pages        = {155--175},
  year         = {2018},
  month        = {},
  publisher    = {Springer},
  organization = {},
  doi          = {}
}

@incollection{Re_Marti_2019,
  author       = {Mart{\'i}, Rafael and Aceves, Ricardo and Le{\'o}n, Maria Teresa and Moreno-Vega, Jose M. and Duarte, Abraham},
  title        = {Intelligent Multi-Start Methods},
  journal      = {},
  booktitle    = {Handbook of Metaheuristics},
  volume       = {},
  number       = {},
  pages        = {221--243},
  year         = {2018},
  month        = {},
  publisher    = {Springer},
  organization = {},
  doi          = {}
}

@article{Re_Hu_2020,
  author       = {Hu, Kai-Cheng and Tsai, Chun-Wei and Chiang, Ming-Chao},
  title        = {A Multiple-Search Multi-Start Framework for Metaheuristics for Clustering Problems},
  journal      = {IEEE Access},
  booktitle    = {},
  volume       = {8},
  number       = {},
  pages        = {96173--96183},
  year         = {2020},
  month        = {},
  publisher    = {IEEE},
  organization = {},
  doi          = {}
}

@article{dokmanic2015euclidean,
  author       = {Dokmanic, Ivan and Parhizkar, Reza and Ranieri, Juri and Vetterli, Martin},
  title        = {Euclidean Distance Matrices: Essential Theory, Algorithms, and Applications},
  journal      = {IEEE Signal Process. Mag.},
  booktitle    = {},
  volume       = {32},
  number       = {6},
  pages        = {12--30},
  year         = {2015},
  month        = {},
  publisher    = {IEEE},
  organization = {},
  doi          = {}
}

@book{borg2005modern,
  author       = {Borg, Ingwer and Groenen, Patrick J. F.},
  title        = {Modern Multidimensional Scaling: Theory and Applications},
  journal      = {},
  booktitle    = {},
  volume       = {},
  number       = {},
  pages        = {},
  year         = {2005},
  month        = {},
  publisher    = {Springer},
  organization = {},
  doi          = {}
}

@article{so2007theory,
  author       = {So, Anthony Man-Cho and Ye, Yin Yu},
  title        = {Theory of Semidefinite Programming for Sensor Network Localization},
  journal      = {Math. Program.},
  booktitle    = {},
  volume       = {109},
  number       = {2},
  pages        = {367--384},
  year         = {2007},
  month        = {Mar.},
  publisher    = {Springer},
  organization = {},
  doi          = {}
}

@article{li2017inexact,
  title={An inexact smoothing {Newton} method for {Euclidean} distance matrix optimization under ordinal constraints},
  author={Li, Qing Na and Qi, Hou-Duo},
  journal={J. Comput. Math.},
  volume       = {35},
  number       = {4},
  pages={469--485},
  year={2017},
  publisher={JSTOR}
}

@article{lu2020feasibility,
  title={Feasibility and a fast algorithm for {Euclidean} distance matrix optimization with ordinal constraints},
  author={Lu, Si Tong and Zhang, Miao and Li, Qing Na},
  journal={Comput. Optim. Appl.},
  volume={76},
  number={2},
  pages={535--569},
  year={2020},
  publisher={Springer},
  doi = {10.1007/s10589-020-00189-9}
}

@article{zhang2020multistatic,
  title={Multistatic Moving Object Localization by a Moving Transmitter of Unknown Location and Offset}, 
  author={Zhang, Yang and Ho, K. C.},
  journal={IEEE Trans. Signal Process.}, 
  booktitle    = {},
  volume={68},
  number={},
  pages={709--728},
  year         = {2020},
  month        = {July},
  publisher    = {IEEE},
  organization = {},
  doi          = {10.1109/TSP.2020.3008752}
}

@article{wu2025multistatic,
  author       = {Wu, Lie Hu and Qin, Guo Dong and Chen, Duo Fang and You, Min Yi and Zou, Yan Bin},
  title        = {Multistatic Target Localization Exploiting Multiple Transmitters with Imperfect Time Synchronization},
  journal      = {IEEE Trans. Aerosp. Electron. Syst.},
  booktitle    = {},
  volume       = {32},
  number       = {6},
  pages        = {1--10},
  year         = {2025},
  month        = {},
  publisher    = {IEEE},
  organization = {},
  doi          = {10.1109/TAES.2025.3605903}
}

@article{larsson2025single,
  author       = {Larsson, Martin and Larsson, Viktor and {\AA}str{\"o}m, Kalle and Oskarsson, Magnus},
  title        = {Single-source localization as an eigenvalue problem},
  journal      = {IEEE Trans. Signal Process.},
  booktitle    = {},
  volume       = {73},
  number       = {},
  pages        = {574-583},
  year         = {2025},
  month        = {},
  publisher    = {IEEE},
  organization = {},
  doi          = {10.1109/TSP.2025.3532102}
}

@article{Varshney2026,
  author  = {Varshney, Piyush and Panwar, Kuntal and Babu, Prabhu},
  title   = {A majorization minimization approach for {AUV}-aided target localization using doppler-shift sensor measurements},
  journal = {Digit. Signal Process.},
  volume  = {168},
  pages   = {105489},
  year    = {2026}
}

@article{Geng2025,
  author  = {Geng, Jun and Zhang, Xun and Guo, Yi Jia and Li, Hao Ran},
  title   = {3-{D} target localization using {BR} and {SA} measurements: closed-form solution and performance analysis},
  journal = {Digit. Signal Process.},
  volume  = {166},
  pages   = {105352},
  year    = {2025}
}

@article{Xu2025,
  author  = {Xu, Shao Hong and Yang, Ming Hai and Li, Cheng Yu and Tang, Bei Chuan and Yang, Yan Bing and Chen, Liang Yin and Sun, Yi Mao},
  title   = {Joint source localization and propagation speed estimation using {TDOA} with hypothesized propagation speed},
  journal = {Digit. Signal Process.},
  volume  = {159},
  pages   = {104934},
  year    = {2025}
}

@article{Sun2022,
  author  = {Sun, Ting and Yu, Ze-Hua},
  title   = {Moving target localization in distributed {MIMO} radar systems with sensor position errors in the presence of a calibration object},
  journal = {Digit. Signal Process.},
  volume  = {131},
  pages   = {103751},
  year    = {2022}
}

@article{bai2016tackling,
  author  = {Bai, Shuang Hua and Qi, Hou-Duo},
  title   = {Tackling the flip ambiguity in wireless sensor network localization and beyond},
  journal = {Digit. Signal Process.},
  volume  = {55},
  pages   = {85--97},
  year    = {2016}
}

@article{zeng2017outlier,
  title={Outlier-Robust Matrix Completion via 
 {$\ell_p$}-Minimization},
  author={Zeng, Wen-Jun and So, Hing Cheung},
  journal={IEEE Trans. Signal Process.},
  volume={66},
  number={5},
  pages={1125--1140},
  year={2018},
  publisher={IEEE}
}

@article{li2022fast,
  title={Fast robust matrix completion via entry-wise {$\ell_0$}-norm minimization},
  author={Li, Xiao Peng and Shi, Zhang-Lei and Liu, Qi and So, Hing Cheung},
  journal={IEEE Trans. Cybern.},
  volume={53},
  number={11},
  pages={7199--7212},
  year={2023},
  publisher={IEEE}
}

@article{wang2025robust,
  title={Robust low-rank matrix completion via sparsity-inducing regularizer},
  author={Wang, Zhi-Yong and So, Hing Cheung and Zoubir, Abdelhak M.},
  journal={Signal Process.},
  volume={226},
  pages={109666},
  year={2025},
  publisher={Elsevier}
}

@article{cambier2016robust,
  title={Robust Low-Rank Matrix Completion by {Riemannian} Optimization},
  author={Cambier, L{\'e}opold and Absil, P.-A.},
  journal={SIAM J. Sci. Comput.},
  volume={38},
  number={5},
  pages={S440--S460},
  year={2016},
  publisher={SIAM}
}

@article{zhang2019localization,
  title={Localization From Incomplete {Euclidean} Distance Matrix: Performance Analysis for the {SVD-MDS} Approach},
  author={Zhang, Huan and Liu, Yulong and Lei, Hong},
  journal={IEEE Trans. Signal Process.},
  volume={67},
  number={8},
  pages={2196--2209},
  year={2019},
  publisher={IEEE}
}






\end{document}